\documentstyle[referee]{mn}
\input{psfig.sty}
\def \be{\begin{eqnarray}}
\def \ee{\end{eqnarray}}
\newcommand{\beq}{\begin{eqnarray}}
\newcommand{\eeq}{\end{eqnarray}}
\newcommand{\beqn}{\begin{eqnarray*}}
\newcommand{\eeqn}{\end{eqnarray*}}

\def\msun{M_\odot}
\begin{document}

\title[Gravitational waves from tidal interactions]
{Gravitational signals due to tidal interactions between white dwarfs and black holes}
\author[C. Casalvieri, V. Ferrari, A. Stavridis]
{C. Casalvieri$^1$, V. Ferrari$^{1}$, A. Stavridis$^2$\\
$^1$ Dipartimento di Fisica ``G.Marconi",
 Universit\` a di Roma ``La Sapienza"\\
 and Sezione INFN  ROMA1, piazzale Aldo  Moro
 2, I-00185 Roma, Italy \\
$^2$ Department of Physics, Aristotle University of Thessaloniki,
 Thessaloniki 54006, Greece
}

\maketitle

\begin{abstract}
In this paper we compute the gravitational signal emitted when a white dwarf
moves around a black hole on a closed or open orbit using the affine model approach. 
We compare  the orbital and the tidal contributions to the signal, 
assuming that the star moves in a safe region where, although very close to the black hole, 
the strength of the tidal interaction is insufficient to provoque the stellar disruption.
We show that for all considered orbits the tidal signal presents sharp peaks
corresponding to the excitation of the star non radial oscillation modes, the amplitude of
which  depends on how deep the star penetrates the black hole tidal radius
and on the type of orbit. Further structure is added to the emitted signal by the 
coupling between the orbital and the tidal motion.
\end{abstract}

\begin{keywords}
gravitational waves --- stars: oscillations --- relativity --- stars:white dwarfs
\end{keywords}

%%%%%%%%%%%%%%%%%%%%%%%%%%%%%%%%%%%%%%%%%%%%%%%%%%%%%%%%%%%%%%%%%%%%%%%%%%%%%
\section{Introduction}

Recent astronomical observations have shown evidence of interactive
processes in action between stars and black holes.  The ESO Very Large
Telescope (VLT) has observed stars that move on very elliptic orbits
around the supermassive black hole  at the center of our Galaxy \cite{VLT}, and
X-ray satellite CHANDRA has monitored
the event of a star disrupted and swollen by the giant black hole 
at the center of the galaxy RXJ1242-11 \cite{chandra_exp}. These and similar phenomena that take
place around black holes generate gravitational waves essentially because of
two mechanisms: one related to the time variation of
the quadrupole moment of the system star-black hole due to the orbital motion, the second to
the quadrupole moment of the star which changes in time because the star is deformed by the
tidal interaction with the black hole.  
The orbital contribution has been computed  using the Post-Newtonian
formalisms
(see for instance \cite{blanchet_Living_Reviews} and references therein).
In this paper we shall focus on the second mechanism: we shall consider
a white dwarf (WD) moving on  assigned orbits around a
black hole (BH), we shall compute how the stellar shape and structure change
along its motion and the gravitational wave (GW) 
signal emitted due to this time-varying deformation.

The problem of tidal deformation has been widely investigated  both in newtonian gravity
and in general relativity, using different approaches.

In 1977 Press and Teukolsky studied the formation of binary systems by two-body tidal
capture using a perturbative approach and provided estimates of the amount of orbital energy 
absorbed in the encounter \cite{press_teukolski}.

In the early eighthies a formalism was  developed 
\cite{carter_luminet_1982,carter_luminet_1983,carter_luminet_1985}
that allows to compute the tidal deformation by integrating a set of equations 
which describe the motion of an element of stellar fluid due to the effect of the 
tidal tensor of the other massive body, on the assumption that while deforming the star
maintains an ellipsoidal form (affine model).  This approach, initially developed in the
framework of Newtonian gravity, was subsequently generalized  to general relativity
when one of the interacting bodies is a Schwarzschild or a Kerr black hole
\cite{marck_1983,luminet_marck_1985}.
The affine model has been used in several papers addressing  some relevant 
astrophysical problems: in
\cite{luminet_carter_1986,luminet_pichon_1989_a,Bicknell_Gingold_1983}
the authors investigated whether, in the 
interaction with a massive companion, a main sequence (MS) star can be compressed  
to such an extent that
the temperature increase in the core  can ignite a detonation of the nuclear fuel 
(a process which may occur in active galactic nuclei having a central massive black hole);
the tidal  capture of a main sequence star and its disruption have been investigated
in \cite{carter_1992} and in \cite{kochanek_1992_a}, respectively; 
tidal processes occurring in neutron star-neutron star coalescing binaries before contact
has been investigated in \cite{kochanek_1992_b}, where
the phase shift  induced on the emitted gravitational signal has also been computed;
the chemical and mechanical behaviour of a 
WD passing inside the tidal radius of a black hole has been studied in
 \cite{luminet_pichon_1989_b}; furthermore, 
the affine model has also been used in \cite{kosovichev_novikov_1992,diener_etal_1995} to study the
tidal capture and interaction of a star and a massive BH.

Tidal effects in close compact binaries  have also been studied using different
formalisms; for instance,  by integrating the hydrodynamical equations in
3D, in a form adapted to express the evolution of the
star principal axes and of other relevant quantities, 
and including radiation reaction and viscous
dissipation \cite{lai_rasio_shapiro_1993,lai_rasio_shapiro_1994_a,lai_rasio_shapiro_1994_b,rasio_shapiro_1995,lai_shapiro_1995}. 
An alternative approach to the integration of the 3D-hydrodynamical
equations has been used in \cite{frolov_etal_1994} to study disruptive and non disruptive
encounters between   white dwarfs and  black holes, focusing on the determination of the periastron
distance, time delay, relativistic precession and tidal forces.

Further studies on tidal interactions in close binaries  are in
\cite{evans_kochanek_1989,novikov_pethick_polnarev_1992,laguna_etal_1993,khokhlov_novikov_pethick_1993_a,khokhlov_novikov_pethick_1993_b,ogawaguchi_kojima_1996,marck_lioure_bonazzola_1996,diener_etal_1997,fryer_etal_1999,ferrari_dandrea_berti_2000,shibata_uryu_2001,gualtieri_etal_2001,berti_ferrari_2001,ivanov_novikov_2001,pons_2002,ivanov_cernyakova_novikov_2003,gomboc_cadez_2005}.

In this paper we use the Carter-Luminet approach to study the tidal interaction 
of a white dwarf and a black hole, with the purpose of computing the 
gravitational signal emitted in this process.
We shall integrate the Carter-Luminet equations for  open and closed orbits 
choosing a white dwarf mass $M_* =1 ~\msun$ and a black hole mass $M_{BH}= 10~\msun$. 
The  WD will be modeled using a  polytropic equation of state (EOS) 
with adiabatic index $\gamma= 5/3$.
The gravitational radiation will be computed by using the quadrupole approach 
and the orbital and the tidal contributions will be compared.
We shall further discuss how the results change for higher values of the black hole mass.

Our study is  based on several simplifying assumptions:\\
- The WD mass is assumed to be  much smaller than the black hole mass. \\
- The black hole does not rotate and the equations of motion of the star,
its internal structure and the black hole tidal tensor are computed using the
equations of newtonian gravity; thus, we call ``black hole'' an object which is
basically a point mass star.  We plan to extend our calculations to general relativity
and to rotating black holes in the near future.
\\
- We neglect tidal effects on the orbital motion.\\
-  We neglect viscous effects, an hypothesis which appears to be justified  in the case of
WD-BH binaries \cite{wiggins_lai_apj_2000}.

The plan of the paper is the following. In section~ 2 we summarize the equations 
we use to compute the tidal deformation of the star in the affine model
approach, and those needed to compute the emitted radiation. 
In section~ 3 we discuss the results of our study: 
for the $1~\msun- 10~\msun$ WD-BH binary we compare the orbital and the
tidal contributions to the gravitational signal, for circular, elliptic and parabolic 
orbits;  we further  discuss how these results depend  on the black hole mass.
Conclusions are drawn in section~ 4.

%%%%%%%%%%%%%%%%%%%%%%%%%%%%%%%%%%%%%%%%%%%%%%%%%%%%%%%%%%%%%%%%%%%%%%%%%%%%%%%%
\section{The relevant  equations}
%%%%%%%%%%%%%%%%%%%%%%%%%%%%%%%%%%%%%%%%%%%%%%%%%%%%%%%%%%%%%%%%%%%%%%%%%%%%%%%%
As mentioned in the introduction, the equations we use to
describe the tidal deformation of the star are fully Newtonian,
and have been developed in  \cite{carter_luminet_1982,carter_luminet_1983,carter_luminet_1985}. 

Let us consider a spherical star with mass $ M_{*} $ and radius $ R_{*} $ in equilibrium,
and be ${\hat r}$ the 
Cartesian position vector of the generic fluid element with respect to a
frame relative to the center of mass of the star in this unperturbed configuration. 
To hereafter, the `hat'  will indicate quantities computed in this frame 
for the unperturbed star.

We shall assume that the star is composed of a perfect fluid with a polytropic EOS
\beq
\label{politropica_gamma}
P= K\rho^{\gamma}.
\eeq
$M_{*}$ will move on  assigned orbits around a non rotating, massive body
$ M_{BH} \gg M_{*} $, to which we associate a reference frame
${\cal{O}}_{BH}$.   Be  $\vec{X}(t)$ the vector which identifies
the position of the center of mass of the star
with respect to this frame.

While moving around $M_{BH}$, the star is deformed by the tidal interaction, and
we shall  indicate as  ${\cal{O}}_*$ the reference frame relative 
to the center of mass of the moving  star
in a parallel propagated frame; in this frame, the position of each fluid element will be
given by the vector  $ \vec r(t)$.  Thus, the position of 
the generic  fluid element   of the star with respect to ${\cal{O}}_{BH}$ will be
\beq
\label{coordinate}
x_i=X_i+r_i,
\eeq
and $ \vec r$ will be a function of time, related to ${\hat r}$ by 
\beq
{\vec r}_i=q_{ij}{\hat r}_j \;\;\;\;\;    (i,\,j=1,\,2,\,3)
\label{affine}
\eeq
where  $ q_{ij}$ is a $3 \times 3$ uniform matrix, i.e.  it depends on time 
but it is independent of {\itshape r}, which implies that the star can assume 
only {\itshape ellipsoidal} configurations (affine model).
$ q_{ij}$ is the matrix we need to compute to know how the stellar shape
changes along the orbit.

Under the assumption of polytropic EOS, the  $q_{ij}$ satisfy the following 
equations \cite{carter_luminet_1983}
\beq
\label{eqmotopolitrop}
\ddot{q_{ij}}=C_{ik}q_{kj}+\frac{\Pi_*}{\mathcal M_*}
\left[
\frac{\Pi}{\Pi_*}q_{ji}^{-1}-\frac{3}{2}\int_0^{\infty}
{du\frac{(S+uI)_{ni}^{-1}}{\Delta}q_{nj}}
\right],
\eeq
where
\beq
\label{tidal_tensor_newton}
C_{ij}=-\frac{\partial^2\Phi_E}{\partial{X_i}\partial{X_j}}
\eeq
is the black hole tidal tensor computed from the Newtonian potential 
\be
\label{newt_pot}
\Phi_E=\frac{G M_{BH}}{X}.
\ee
In eqs. (\ref{eqmotopolitrop})
$\Pi$ is the volume integral of the local pressure
\be
\label{defPI}
\Pi &\equiv& \int{P\cdot dV},
\ee
and, since 
\be
\label{scaledens}
\rho={\hat \rho}\cdot||q||^{-1},
\ee
where ${\hat \rho}$ is the density in the spherical unperturbed configuration
and $||q|| \equiv \det ~q$, 
it can be written as 
\be
\label{Pi}
\Pi~=~K\int{\rho^{\gamma}dV} ~=~K||q||^{1-\gamma}\int{\hat{\rho}^{\gamma-1}dM}
~=~||q||^{1-\gamma}\int{\hat{P}dV}
~\equiv~ \Pi_*~||q||^{1-\gamma}.
\eeq
$\Pi_*$ is the value of $\Pi$ in the  unperturbed configuration.
Furthermore,
${\mathcal{M_*}}$ is the scalar quadrupole moment of the initial configuration
\be
\label{scalare_quadrupolo}
{\mathcal{M_*}}=\frac{1}{3}\int{\hat{r}_i\hat{r}_idM}.
\ee
The function $\Delta$ in eq. (\ref{eqmotopolitrop}) is 
\be
\Delta=\sqrt{||S+uI||},
\ee
where $I$ is the identity matrix,
\beq
\label{S}
S_{ij}=q_{ik}q_{jk},
\eeq
and $u$ is the integration variable of the integral appearing in 
(\ref{eqmotopolitrop}).

If the star is initially non rotating (as we shall assume), or if its initial spin is
orthogonal to the orbital plane, the non vanishing components of $q_{ij}$ 
reduce to  \cite{carter_luminet_1983}
\beq
\label{q_blocchi}
q_{ij}(t) = \left(
\begin{array}{ccc}
q_{11} & q_{12} & 0 \\
q_{21} & q_{22} & 0 \\
0 & 0 & q_{33}
\end{array}
\right).
\eeq

%%%%%%%%%%%%%%%%%%%%%%%%%%%%%%%%%%%%%%%%%%%%%%%%%%%%%%%%%%%%%%%%%%%%%
\subsection{Gravitational radiation}
%%%%%%%%%%%%%%%%%%%%%%%%%%%%%%%%%%%%%%%%%%%%%%%%%%%%%%%%%%%%%%%%%%%%%
The gravitational signal emitted by the system  will be computed
by using the quadrupole formalism, according to which 
\beq
\label{h_di_emmeTT}
\left\{
\begin{array}{l}
h_{\mu 0}^{TT}=0\;\;\;\;\; \mu=0,\,3\\
h_{ik}^{TT}(t,r)=\frac{2G}{c^4r}\left[\frac{d^2}{dt^2}m_{ik}^{TT}
\left(t-\frac{r}{c}\right)\right]
\end{array}
\right.
\eeq
where $m_{ij}^{TT}$ is the source quadrupole moment projected 
on the {\bf TT}-gauge
\beq
\label{quadrupoloTT}
m_{ij}^{TT}={\mathcal {P}}_{ijlm}m_{lm}
\eeq
and ${\mathcal {P}}_{ijlm}$ is the projector tensor.
%%%%%%%%%%%%%%%%%%%%%%%%%%%%%%%%%%%%%%%%%%%%%%%%%%%%%%%%%%%
For the system under consideration, the quadrupole moment can be evaluated as follows
\be
\label{mom_quad1}
m_{ik}(t)=\frac{1}{c^2}\int_V{T^{00}\left(t,x_j\right)x_ix_kd^3x}\equiv
\int_V{\rho\left(t,r_j\right)\left(X_i+r_i\right)\left(X_k+r_k\right)d^3r},
\ee
i.e.
\beq
\label{mom_quad2}
m_{ik}(t)&=&\int_V{\rho\left(t,r_j\right)X_i(t)X_k(t)d^3r}+
\int_V{\rho\left(t,r_j\right)r_ir_kd^3r}\nonumber\\
&+&X_i(t)\int_V{\rho \left(t,r_j\right)r_kd^3r}+X_k(t)\int_V{\rho\left(t,r_j\right)r_id^3r},
\eeq
where we remind that the $X_i(t)$ are the coordinates of the center of mass of the star
with respect to ${\cal{O}}_{BH}$,
whereas $r_k(t)$ are the coordinates of the element of stellar fluid with respect to 
${\cal{O}}_{*}$.
The first integral is the standard quadrupole moment associated to the orbital motion
\beq
\label{mom_quad_orb}
m_{ik}^{orb}(t)=X_i(t)X_k(t)\int_V{\rho\left(t,r_j\right)d^3r}=M_*X_i(t)X_k(t).
\eeq
The second integral is the contribution due to the tidal interaction
which, using eqs.  (\ref{affine}) can be written as
\beq
\label{mom_quad_espl3}
m_{ik}^{def}(t)=\int_V{\rho\left(t,r_j\right)r_ir_kd^3r}=q_{il}(t)q_{km}(t)
\int_V{{\hat \rho}\left(t,{\hat r}_{j}\right){\hat r}_{l}{\hat r}_{m}d^3{\hat r}}.
\eeq
In the last equality we have used  eq. (\ref{scaledens}) and
the property that the volume scales as
\beq
\label{riscal_volume}
d^3r=||q||d^3{\hat r}.
\eeq
Using eqs.  (\ref{scalare_quadrupolo}) and  (\ref{S}), eq. (\ref{mom_quad_espl3}) becomes 
\beq
\label{mom_quad_def}
m_{ik}^{def}(t)=q_{il}(t)q_{km}(t)\int_V{{\hat r}_{l}{\hat r}_{m}dM}=
  {\mathcal{M_*}}S_{ik}(t).
\eeq
The last two integrals in (\ref{mom_quad1}) vanish due to symmetry; indeed
\beq
X_i(t)\int_V{\rho\left(t,r_j\right)r_kd^3r}=X_i(t)q_{km}(t)\int_V{{\hat \rho}
\left(t,{\hat r}_j\right){\hat r}_{m}d^3{\hat r}}=0
\eeq
since for every fluid element that, in the unperturbed configuration, has coordinates
$r_j=\left(x_j,\,y_j,\,z_j\right)$  there exists one with coordinates
$\left(-x_j,\,-y_j,\,-z_j\right)$.
Thus, in conclusion, the quadrupole moment of the system is composed of two contributions,
the orbital one, $m_{ik}^{orb}$ ( eq. \ref{mom_quad_orb} ), and that due to the tidal interation,
$m_{ik}^{def}$ (eq. \ref{mom_quad_def}).

%%%%%%%%%%%%%%%%%%%%%%%%%%%%%%%%%%%%%%%%%%%%%%%%%%%%%%%%%%%
\section{Results}
%%%%%%%%%%%%%%%%%%%%%%%%%%%%%%%%%%%%%%%%%%%%%%%%%%%%%%%%%%%
Eqs. (\ref{eqmotopolitrop}) have been integrated 
for a system composed of a  white dwarf described by a polytropic EOS with
$\gamma=\frac{5}{3}$ and  a more massive companion of mass $M_{BH}=10~\msun$.
The WD central density of the unperturbed configuration 
is such  that $M_* = 1~\msun$ and $R_* = 6955 ~km$.

We have considered  circular, elliptic and parabolic orbits, and using 
eqs. (\ref{mom_quad_orb} ), (\ref{mom_quad_def}) and (\ref{h_di_emmeTT})
we have computed the emitted gravitational signal.

It is known that if the periastron of the orbit lays within 
a certain critical radial distance from $M_{BH}$ the star can be
destroyed by the tidal interaction. This radius, $R_{R}$, mainly depends on
the composition of the interacting bodies and on the orbit.
According to Newtonian Gravity $R_{R}$ is approximately 
%\footnote{It should be mentioned that the WD starts to transfer matter on the black hole 
%for a radial distance smaller than $R_R$. This critical distance has been estimated in
%\cite{freyer} to be
%\[
%R_{crit} = R_*\frac{0.6 q^{2/3}+ \ln (1+q^{1/3})}{0.49 q^{2/3}}
%\]
%where $q=M_*/M_BH$.
%For the values of $q$ considered in this paper $R_{crit} \simeq 2 R_R$. }
\be
\label{roche_radisu}
R_{R}\approx \left(\frac{M_{BH}}{\bar \rho}\right)^{\frac{1}{3}}
\approx \left(\frac{M_{BH}}{M_*}\right)^{\frac{1}{3}}R_* ,
\ee
where ${\bar\rho}$ is the star average density.
As usual, we define a penetration factor as the ratio
\be
\label{beta}
\beta= \frac{R_R}{R_p},
\ee
where $R_p$ is the distance of closest approach between the two bodies.

%@@@@@@@@@@@@@@@@@@@@@@@@@@@@@@@@@@@@@@@@@@@@@@@@@@@@@@@@@@@@@@@@@@@@@@@@@@@@2

%%%%%%%%%%%%%%%%%%%%%%%%%%%%%%%%%%%%%%%%%%%%%%%%%%%%%%%%%%%%%%%%%%%%%%%%%
%                            TABLE  1
%%%%%%%%%%%%%%%%%%%%%%%%%%%%%%%%%%%%%%%%%%%%%%%%%%%%%%%%%%%%%%%%%%%%%%%%%
\begin{table}
\caption{The limiting value of $\beta$ above which the tidal interaction between 
the white dwarf and the black hole becomes disruptive is given for the 
orbits considered in this paper.
}
\begin{center}
\begin{tabular}{@{}ccc@{}}
%\begin{tabular}{||p{3.08cm}*{3}{c|}|}
\hline
Orbit & e& $\beta_{crit}$
\\\hline
        &     &   \\[1pt]
Circular  & 0 & 0.78\\[0.5ex]
\hline
Parabolic & 1 & 1.05\\[0.5ex]
\hline
Elliptic  & 0.25 & 0.78\\[0.5ex]
\hline
&0.75 & 0.97\\[0.5ex]
\hline
&0.95 & 1.03\\[0.5ex]
\hline
\end{tabular}
\end{center}
\label{table1}
\end{table}
%\clearpage
%%%%%%%%%%%%%%%%%%%%%%%%%%%%%%%%%%%%%%%%%%%%%%%%%%%%%%%%%%%%%%%%%%%%%%%%%

\begin{center}
\begin{figure}
\centerline{\mbox{
\psfig{figure=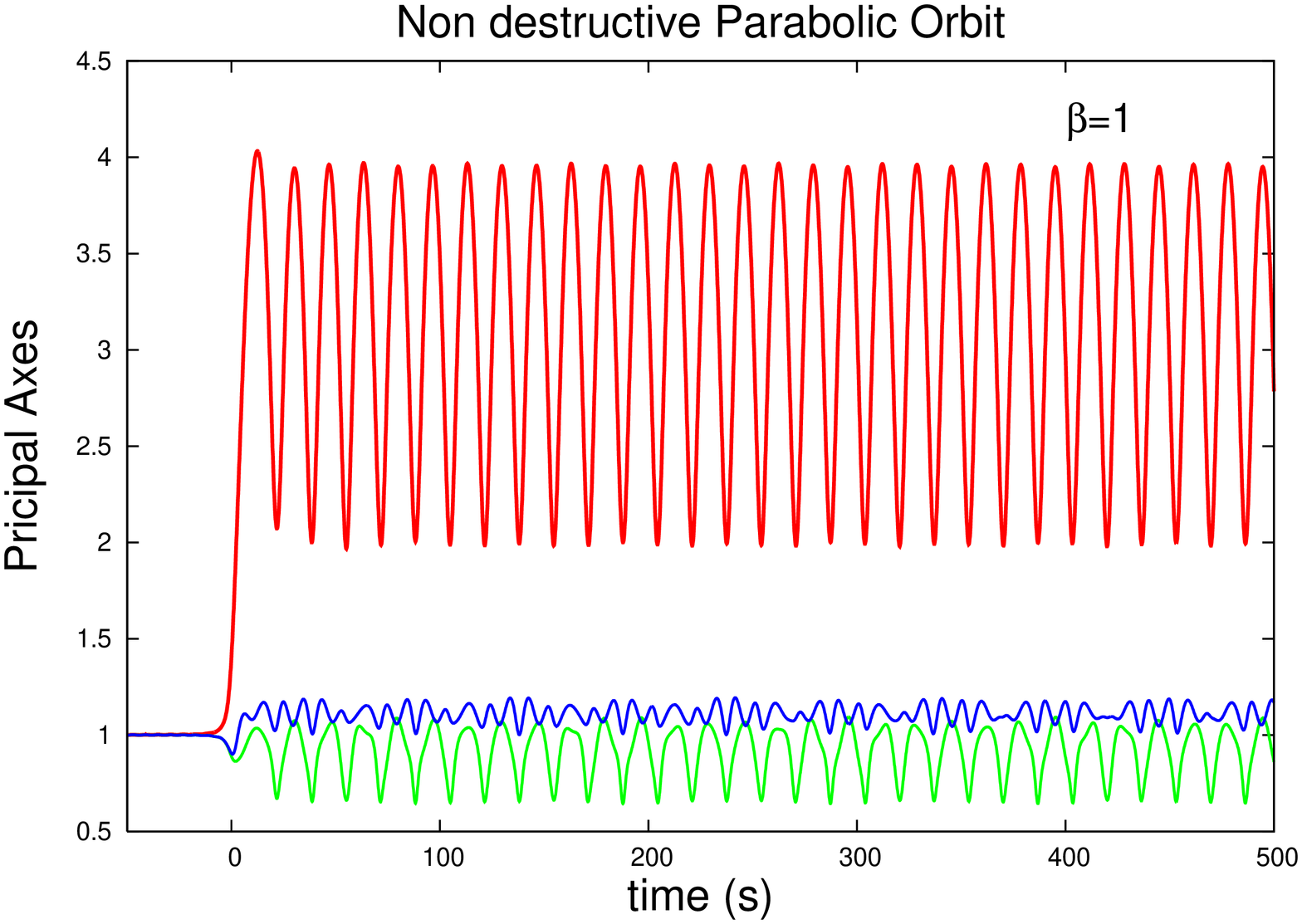,width=9cm,height=7.5cm,angle=-90}
}}
\vskip 12pt
\centerline{\mbox{
\psfig{figure=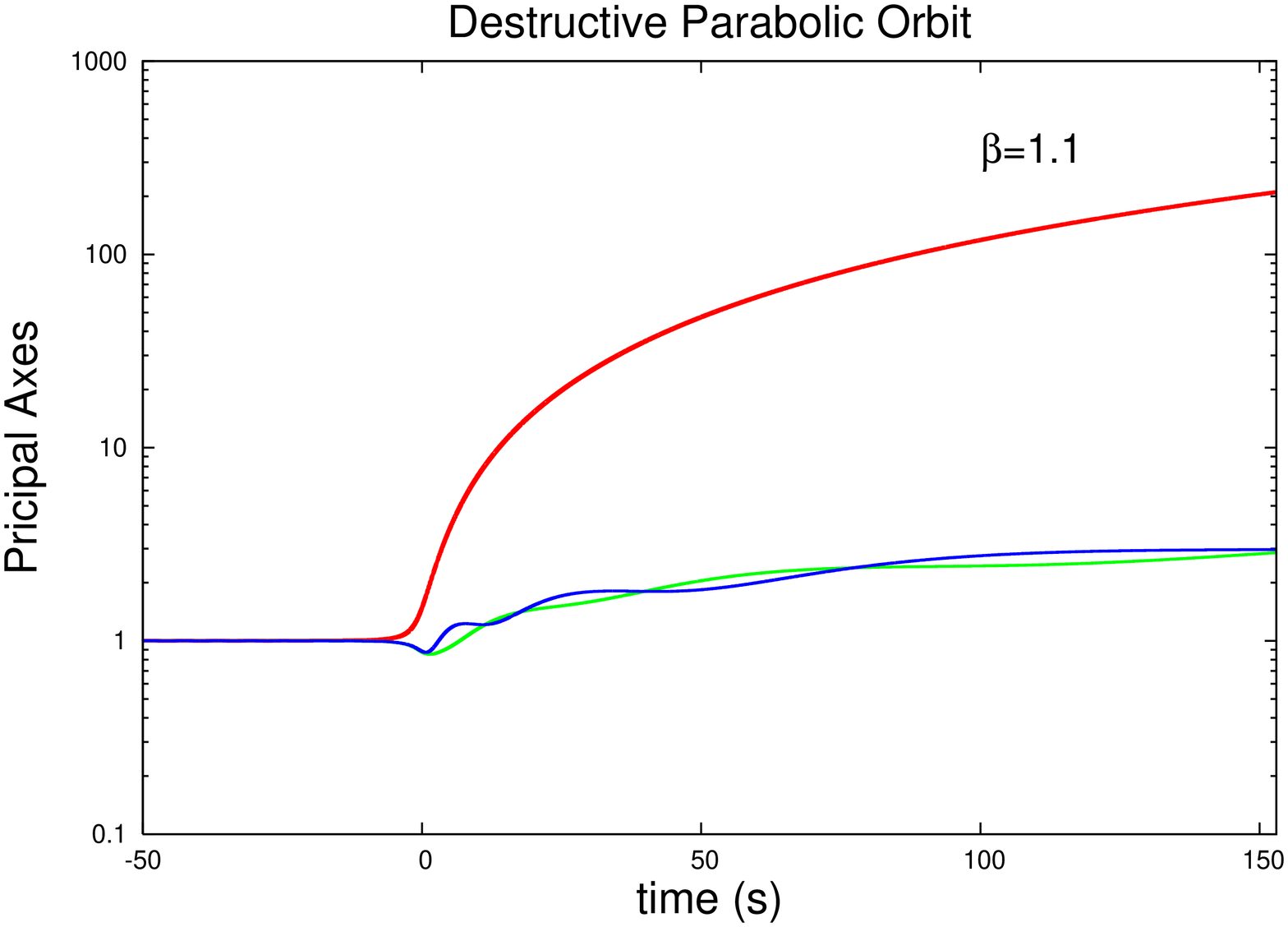,width=9cm,height=7.5cm,angle=-90}
}}
\caption{The white dwarf principal axes are plotted versus time for two  parabolic orbits
with $\beta=1$ and $\beta=1.1$. The indefinite growth of one axis in the lower picture
signals a destructive interaction.}
\label{FIG1}
\end{figure}
\end{center}

In order to establish what is the critical value of $\beta$ below which
the star can safely approach the black hoel without being 
destroyed by the tidal interaction, we proceed as follows.
We monitor the behaviour of the star principal axes as a function
of time for increasing values of $\beta$.
As an example, in figure \ref{FIG1} we plot the axes for two parabolic orbits corresponding to
$\beta=1$ (upper panel) and  $\beta=1.1$ (lower panel log-scale).
In the first case the axes oscillate around average
values that approximately correspond to a conformal Riemaniann stationary configuration
\cite{diener_etal_1995}. 
Conversely, for $\beta=1.1$ we see one axis indefinitely growing, and this signals 
a destructive interaction.
In table \ref{table1} we give the values of $\beta_{crit}$ for circular, elliptic and parabolic
orbits for the system we consider. $\beta_{crit}$ is the limiting value of $\beta$ above which we
observe the indefinite growing of one axis.

In what follows we shall consider only values of $\beta \leq \beta_{crit}$.
In all cases  we shall assume that the orbit lays on the equatorial plane 
($(x,y)$-plane) and we
shall compute the gravitational signal emerging in the $z$-direction in the
{\bf TT}-gauge.
$h^{def}_+$  and $h^{def}_\times$
will refer to the part of the signal due to the tidal interaction
and to the consequent stellar deformation, computed using the quadrupole moment
(\ref{mom_quad_def}),
whereas $h^{orb}_{+}$ and $h^{orb}_{\times}$ will be the orbital 
contribution, computed using eq (\ref{mom_quad_orb}).
%@@@@@@@@@@@@@@@@@@@@@@@@@@@@@@@@@@@@@@@@@@@@@@@@@@@@@@@@@@@@@@@@@@@@@@@@@@@@2

%%%%%%%%%%%%%%%%%%%%%%%%%%%%%%%%%%%%%%%%%%%%%%%%%%%%%%%%%%%%
%                      FIGURE
%%%%%%%%%%%%%%%%%%%%%%%%%%%%%%%%%%%%%%%%%%%%%%%%%%%%%%%%%%%%
%%%%%%%%%%%%%%%%%%%%%%%%%%%%%%%%%%%%%%%%%%%%%%%%%%%%%%%%%%%%

%%%%%%%%%%%%%%%%%%%%%%%%%%%%%%%%%%%%%%%%%%%%%%%%%%%%%%%%%%%
\subsection{Circular orbits}
\begin{center}
\begin{figure}
\centerline{\mbox{
\psfig{figure=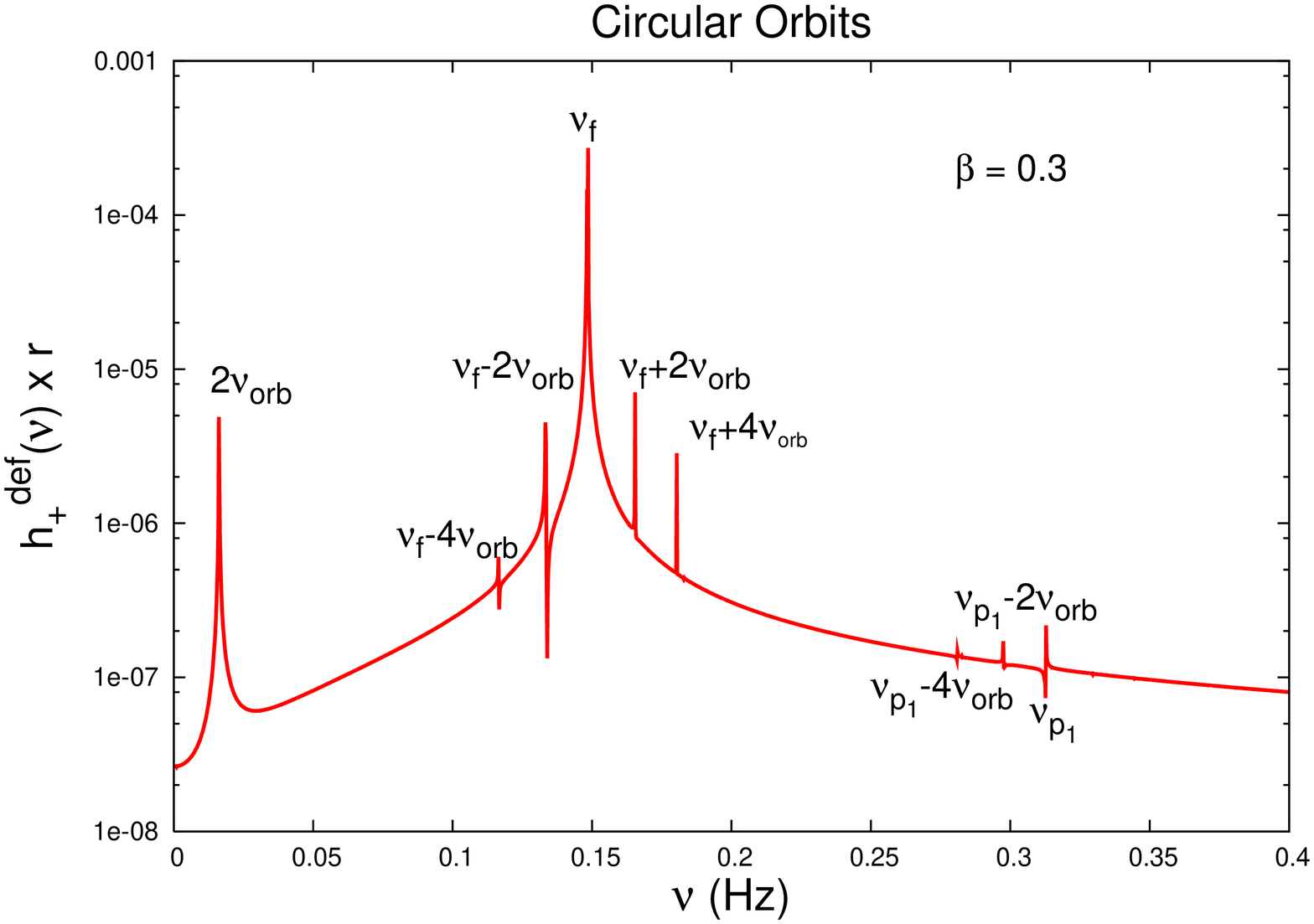,width=9cm,height=7.5cm,angle=-90}
}}
\vskip 12pt
\centerline{\mbox{
\psfig{figure=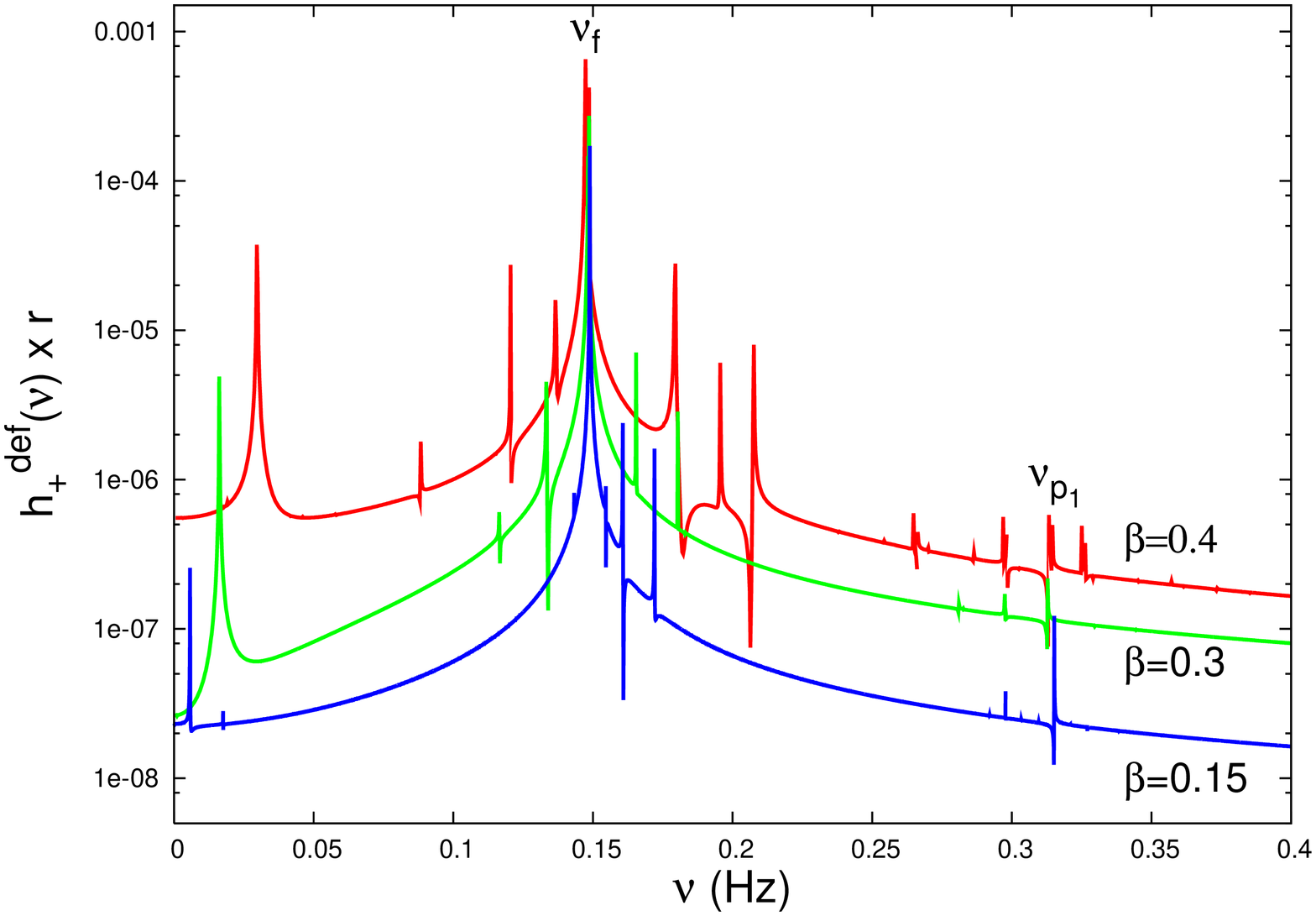,width=9cm,height=7.5cm,angle=-90}
}}
\caption{The Fourier transform of the GW signal associated to the tidal
interaction between a $1~\msun$ WD and a $10~\msun$ BH is plotted as a 
function of frequency for a circular orbit with 
$\beta=0.3$ (up) and for different values of $\beta$ (down).
The main peak  at $\nu_f=0.146$ Hz corresponds to the frequency of 
the fundamental mode of the non radial oscillations  of the star, 
a smaller peak at $\nu=0.316$ Hz indicates the excitation of the 
first {\bf p}-mode.
The remaining harmonics are due to the coupling between orbital and tidal
motion (see text).}
\label{FIG2}
\end{figure}
\end{center}

\begin{center}
\begin{figure}
\centerline{\mbox{
\psfig{figure=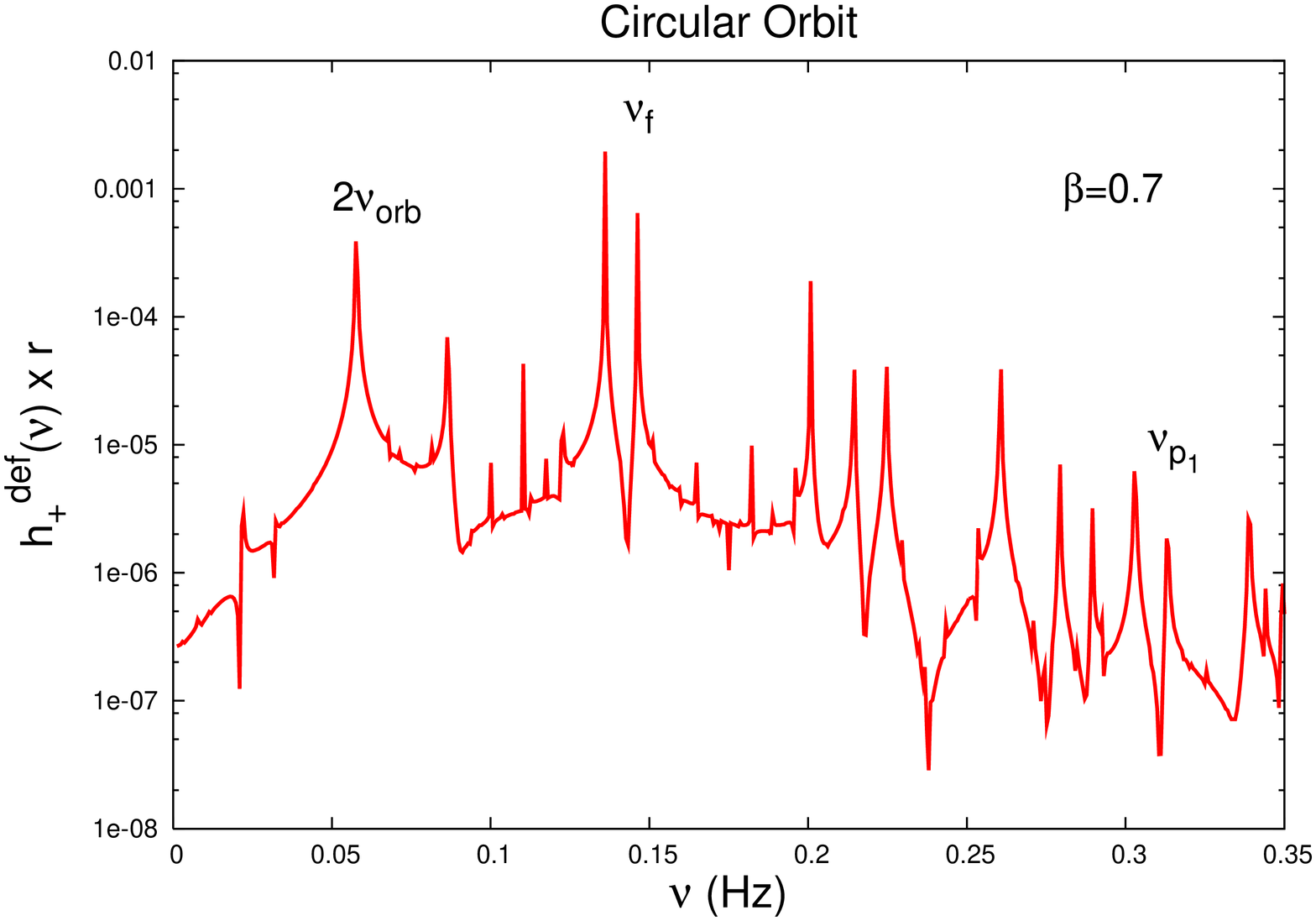,width=9cm,height=7.5cm,angle=-90}
}}
\caption{Tidal signal for a circular orbit with 
$\beta=0.7$. The coupling between orbital and tidal motion is stronger than
in figure \ref{FIG2} (where $\beta$ is smaller) and higher harmonics are excited. }
\label{FIG3}
\end{figure}
\end{center}

%%%%%%%%%%%%%%%%%%%%%%%%%%%%%%%%%%%%%%%%%%%%%%%%%%%%%%%%%%%
When the orbit is circular, the orbital contribution to the gravitational signal is
\be
\label{sigcirc}
h^{orb}_{+}(t)=\imath h^{orb}_{\times}(t)= A\frac{ \cos(2 \omega_{orb} t)}{r},
\ee
where $\omega_{orb}=\left(\frac{GM}{D^3} \right)^{1/2}$ is the Keplerian angular velocity,
$D$ being the separation between the two bodies,
$r$ is the  radial distance from the source, and
\be
\label{amplcirc}
A= \frac{4G^2 M \mu}{c^4}, 
\ee
where $M=M_*+M_{BH}$ is the total mass and $\mu= \frac{M_*~M_{BH}}{M}$ is the reduced mass of the
system.
Thus, radiation is emitted in a single spectral line at a frequency $\nu_{orb}=
\frac{\omega_{orb}}{2\pi}$, i.e.
\be
h^{orb}_{+}(\nu)=\frac{1}{r}~ \frac{A}{4\pi}\left[\delta(\nu-2\nu_{orb}) -
\delta(\nu+2\nu_{orb}  \right].
\ee 
%%%%%%%%%%%%%%%%%%%%%%%%%%%%%%%%%%%%%%%%%%%%%%%%%%%%%%%%%%%%%%%%%%%%%%%%%
%                            TABLE 2
%%%%%%%%%%%%%%%%%%%%%%%%%%%%%%%%%%%%%%%%%%%%%%%%%%%%%%%%%%%%%%%%%%%%%%%%%
\begin{table}
\caption{Orbital parameters and  GW signals amplitudes
are tabulated for different values of the penetration factor $\beta$
(column 1) for circular orbits. 
The orbital radius $R$,  period $T$ and frequency $\nu_{orb}$ are given 
in columns 1 to r43; the adiabatic timescale $\tau$ is in column 5;
in the last two column we give the  amplitude of the spectral line 
of the orbital signal, $A/4\pi$, and that of the {\bf f}-mode peak 
appearing in  the tidal signal; they are both multiplied by the source
distance, $r$.}
\begin{center}
\begin{tabular}{||p{0.08cm}*{7}{c|}|}
\hline
$\beta$ & $R_p$ & T & $\nu_{orb}$ &  $\tau$ & $(A/4\pi)\times r$ & $h_{max}\times r
$\\[0.5ex]
 & $10^4$ (km) & (s) & (Hz) & (s) & $10^{-4}$ (km) & $10^{-4}$  (km)\\[0.5ex]
\hline
0.15 & $14.6$ & $336.4$ & $3.0 \times 10^{-3}$ &
$ 4.8 \times 10^{11} $ & 0.46 & 1.7 \\[0.5ex]
\hline
0.3 & $8.0$ & $118.9$ & $8.5 \times 10^{-3}$ & 
$3.1 \times 10^{10}$ & 0.87 & 2.7 \\[0.5ex]
\hline
0.4 & $6.0$ & $77.2$ & $1.3 \times 10^{-2}$  & 
$1.0 \times 10^{10}$& 1.4 & 4.2\\[0.5ex]
\hline
0.5 & $4.8$ & $55.3$ & $1.8 \times 10^{-2} $ & 
$4.0 \times 10^{9}$ & 1.5 & 7.0 \\[0.5ex]
\hline
0.6 & $4.0$ & $42.0$ & $2.4 \times 10^{-2}$  & 
$1.9 \times 10^{9}$ & 1.7 & 9.2\\[0.5ex]
\hline
0.7 & $3.1$ & $33.4$ & $3.0 \times 10^{-2}$  & 
$1.0 \times 10^{9}$ & 2.5 & 15\\[0.5ex]
\hline
\end{tabular}
\end{center}
\label{table2}
\end{table}
%\clearpage
%%%%%%%%%%%%%%%%%%%%%%%%%%%%%%%%%%%%%%%%%%%%%%%%%%%%%%%%%%%%%%%%%%%%%%%%%
Due to gravitational emission the orbit decays; thus the signal we compute using eqs.
(\ref{sigcirc}) may not be correct, unless the changes induced on the orbit by
the energy lost in GW occur on a timescale
much longer than the orbital period, so that we can neglect radiation reaction effects. 
To check  whether this is the case for the orbits we consider,
we have computed the adiabatic timescale
$\tau$ 
\be
\tau \sim \frac{E_{orb}}{L_{GW}},\ee
where $E_{orb}= -\frac{1}{2}\frac{G M \mu}{D}$ is the system orbital energy, and 
$L_{GW}\equiv \frac{dE_{GW}}{dt}$ is the GW-luminosity  due to the time variation of the
orbital quadrupole moment
\[
L_{GW}= \frac{32~G^4~\mu^2 M^3}{5 ~c^5~D^5}.
\] 
The results of our calculations on the tidal interaction 
are summarized in figure \ref{FIG2} and \ref{FIG3} and in table
\ref{table2}.
%%%%%%%%%%%%%%%%%%%%%%%%%%%%%%%%%%%%%%%%%%%%%%%%%%%%%%%%%%%%
%                      FIGURE
%%%%%%%%%%%%%%%%%%%%%%%%%%%%%%%%%%%%%%%%%%%%%%%%%%%%%%%%%%%%
%%%%%%%%%%%%%%%%%%%%%%%%%%%%%%%%%%%%%%%%%%%%%%%%%%%%%%%%%%%%

In the upper panel of figure \ref{FIG2} we  plot the Fourier  transform of 
the tidal signal $h^{def}_+(\nu)$ for $\beta=0.4$. 
The plot  shows a number of very sharp peaks. The first is at $\nu=
2\nu_{orb}$, and it is a signature of the orbital motion on the tidal deformation of the star. 
The highest peak is at a frequency $\nu_f= 0.146$ Hz, which is the frequency of the 
fundamental mode of the non radial oscillations of the star in the unperturbed
configuration (i.e. when it has a spherical form).
This peak is surrounded by equally spaced peaks, at frequencies
$\nu =\nu_f \pm 2n \nu_{orb}, ~n=1,2,...$, due to the coupling of the orbital motion to the deformation.
In addition, we also see smaller peaks at the frequencies of the first pressure mode,
$\nu_{p_1}=0.316$, and further harmonics $\nu =\nu_{p_1} \pm 2n \nu_{orb}$.
This picture clearly shows that the non radial modes of oscillations of the star
are excited in the tidal interaction and their signature appears in the gravitational 
wave spectrum.
The plot for $h^{def}_\times(\nu)$  is entirely similar and will not be shown.

In the lower panel of figure \ref{FIG2} we show $h^{def}_+(\nu)$ for increasing values of
$\beta$ (i.e. for decreasing orbital radii); we see that, as $\beta$
increases and the orbit shrinks,  
the $2\nu_{orb}$- peak shifts toward higher frequencies,
but the peaks  corresponding to the
excitation of the {\bf f}- and {\bf p}-modes, of course, do not move. Moreover, due to che change
of $\nu_{orb}$ the spacing between the harmonics changes as well.

%%%%%%%%%%%%%%%%%%%%%%%%%%%%%%%%%%%%%%%%%%%%%%%%%%%%%%%%%%%%
%                      FIGURES
%%%%%%%%%%%%%%%%%%%%%%%%%%%%%%%%%%%%%%%%%%%%%%%%%%%%%%%%%%%%
%%%%%%%%%%%%%%%%%%%%%%%%%%%%%%%%%%%%%%%%%%%%%%%%%%%%%%%%%%%%%
%%%%%%%%%%%%%%%%%%%%%%%%%%%%%%%%%%%%%%%%%%%%%%%%%%%%%%%%%%%%%
Finally, in figure \ref{FIG3} we show the signal for  the case $\beta=0.7$ which is close to
tidal disruption (see table \ref{table1}). In this case
the contribution of different harmonics is higher, giving rise to a 
richer structure in the frequency spectrum.

In table \ref{table2},  for several values of $\beta$ we show
the orbital radius $R$,  period $T$ and frequency $\nu_{orb}$ (column 1 to 4);
in column 5 we give the adiabatic timescale $\tau$.
These data show that, for the orbits we consider, we  are always in the adiabatic regime
since $\tau >> T$.
In the last two column we give the  amplitude of the spectral line 
of the orbital signal, $A/4\pi$, where $A$ is given in eq. (\ref{amplcirc}), 
and  the amplitude of the {\bf f}-mode peak appearing in the tidal signal $h^{def}_+$.
It is interesting to note that the {\bf f}-mode occurs at frequencies 
higher than $2\nu_{orb}$, and its amplitude is always larger than that of the obital signal.

It is worth mentioning that  the damping time of the fundamental mode is
very large, of the order of $\tau_{GW} \sim 10^6$ seconds for the stars we consider.

%%%%%%%%%%%%%%%%%%%%%%%%%%%%%%%%%%%%%%%%%%%%%%%%%%%%%%%%%%%
\subsection{Elliptic orbits}
%%%%%%%%%%%%%%%%%%%%%%%%%%%%%%%%%%%%%%%%%%%%%%%%%%%%%%%%%%%

\begin{center}
\begin{figure}
\centerline{\mbox{
\psfig{figure=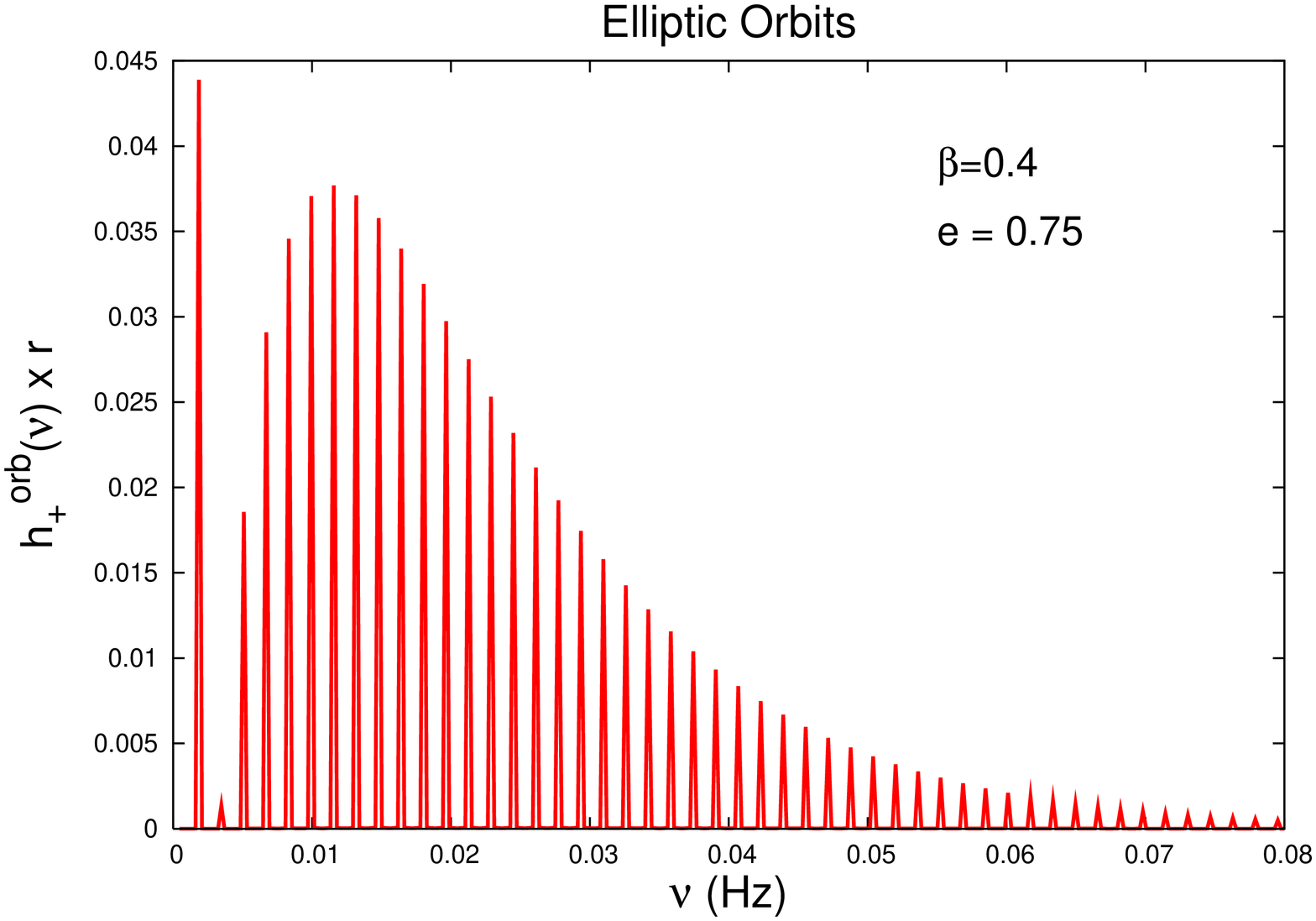,width=9cm,height=7.5cm,angle=-90}~~~~
\psfig{figure=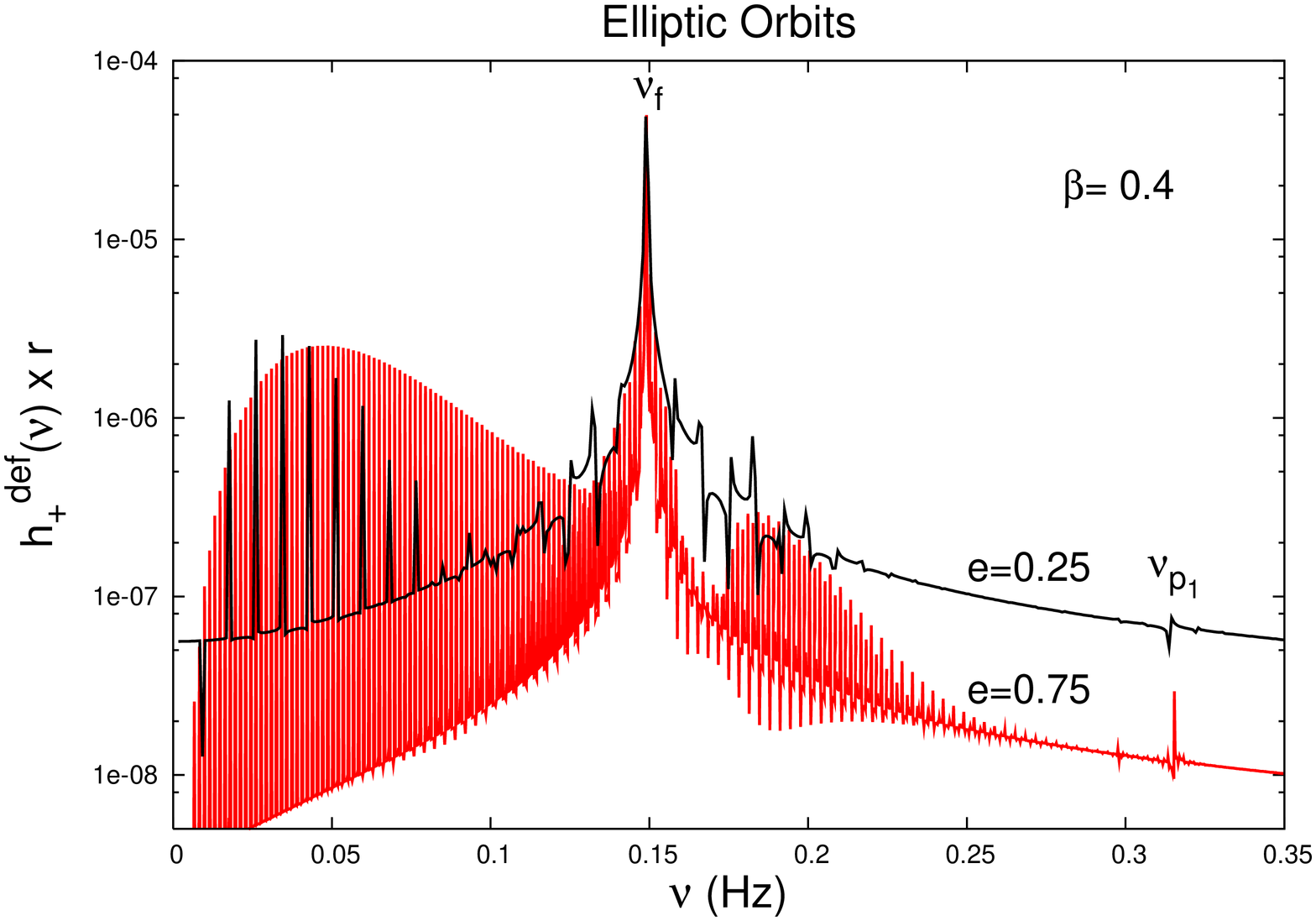,width=9cm,height=7.5cm,angle=-90}
}}
\vskip 12pt
\centerline{\mbox{
\psfig{figure=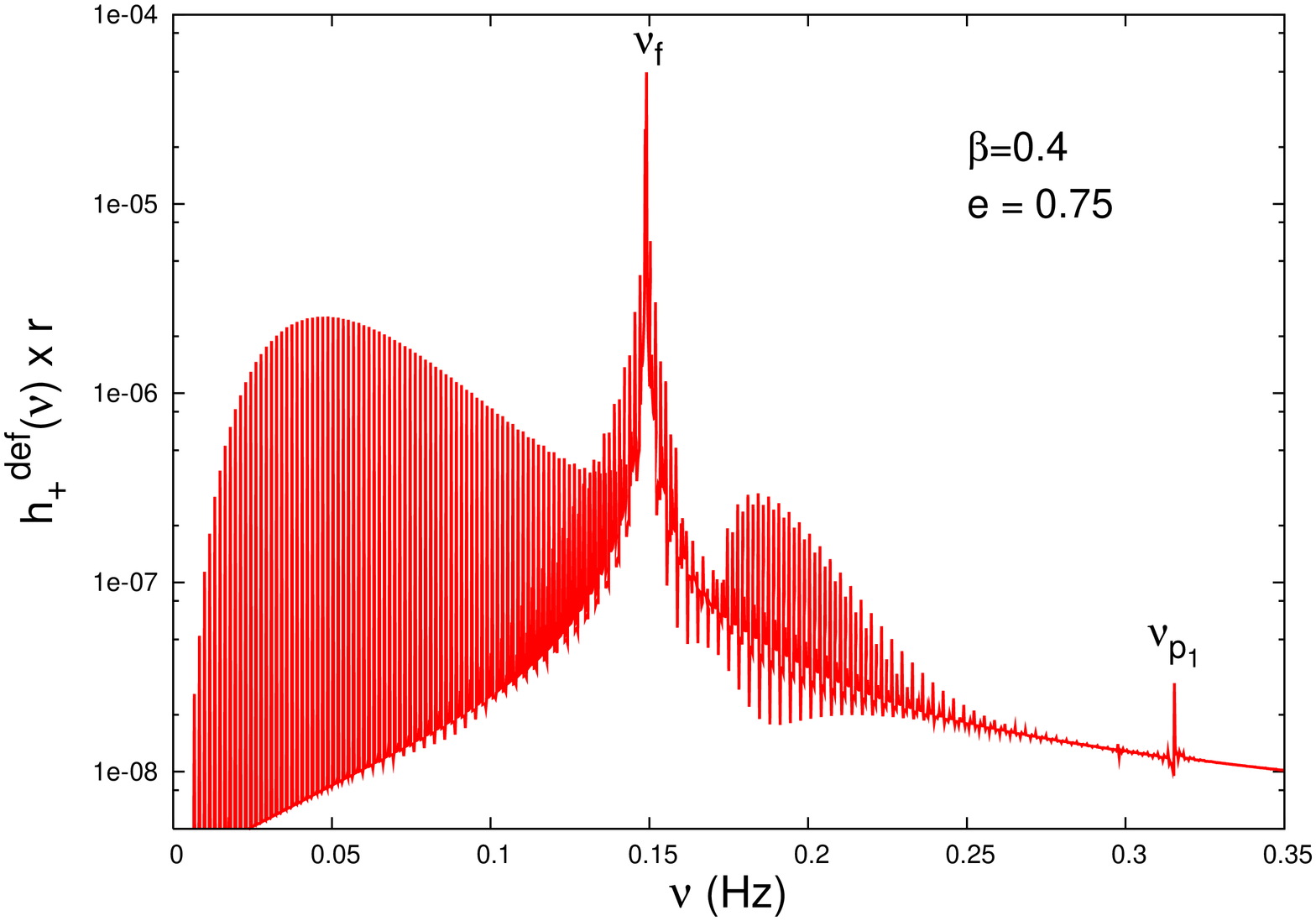,width=9cm,height=7.5cm,angle=-90}~~~~
\psfig{figure=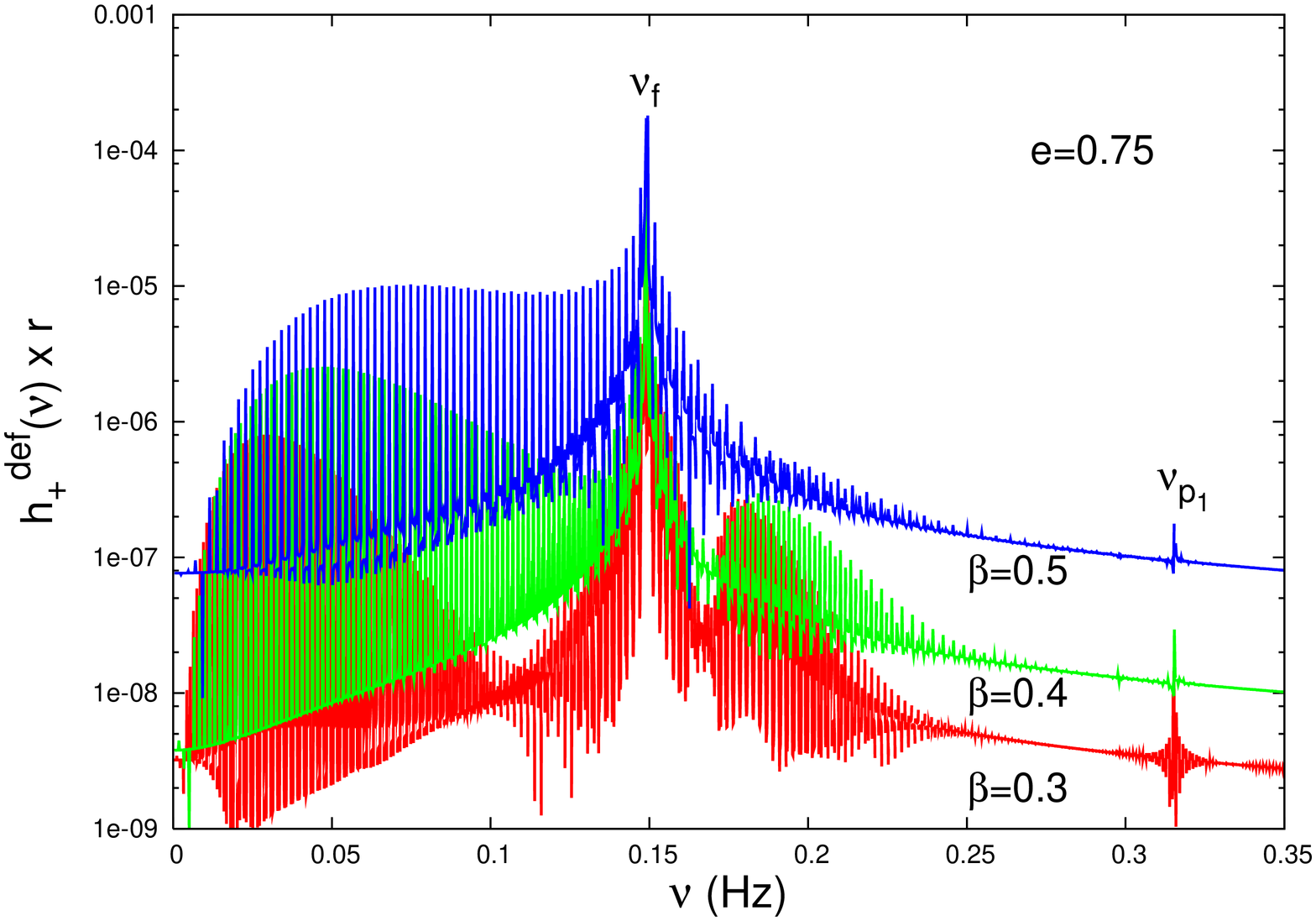,width=9cm,height=7.5cm,angle=-90}
}}
\caption{In the left panel we plot the Fourier transform of the orbital
signal (up) and of the tidal signal (down) emitted by the WD-BH binary
studied in  this paper when the orbit is an ellipse with 
$\beta=0.4$ and $e=0.75$. 
As in the circular case, the main peak in $h_+^{def}$
corresponds to the excitation o the {\bf f}-mode, but its
amplitude is much lower than that of the maximum in $h_+^{orb}$.
In the right panel we show the tidal signal for elliptic orbits with 
$\beta$ assigned and varying $e$ (up), and with $e$ assigned and varying
$\beta$ (down).}
\label{FIG4}
\end{figure}
\end{center}

Whereas for a circular  orbit radiation is  emitted in a single spectral line
at twice the orbital frequency, when the orbit is eccentric waves are
emitted at frequencies multiple of $\nu_{orb}$, and the number of  
equally spaced spectral lines increases with the eccentricity \cite{peters_matews_1963,ferrari_dandrea_berti_2000}.
In the left panel of figure \ref{FIG4} we  compare the orbital (up)
and the tidal (down) signal
emitted on an orbit with $\beta=0.4$ and $e=0.75$.

As in the circular case, the excitation of the {\bf f}- and {\bf p$_1$}-modes 
is manifested by the sharp peaks at the corresponding frequencies, and it is interesting to see
that the coupling between the orbital and tidal motion  introduces a large number of harmonics.
Note also that the maximum of the orbital signal is much larger 
than the {\bf f}-mode peak in tidal signal.

In the right panel of figure \ref{FIG4} we show how the structure of the tidal 
signal changes due to eccentricity, plotting $h^{def}_+(\nu)$ for  $\beta=0.4$ 
and for two values of $e$, (upper panel): as expected,  we see that lower 
eccentricities correspond to a smaller number of
spectral lines due to the orbital-tidal coupling. 
In the same figure (lower panel) we compare the tidal signal emitted on
orbits with the same eccentricty ($e=0.75$) and different  values of $\beta$,
showing how the mode excitation is  sensitive to  the penetration parameter.

We find that for the orbital parameters tabulated in table \ref{table3}, the
maximum of the orbital signal is always much larger than the {\bf f}-mode peak in tidal signal.
%%%%%%%%%%%%%%%%%%%%%%%%%%%%%%%%%%%%%%%%%%%%%%%%%%%%%%%%%%%%
%                      FIGURE
%%%%%%%%%%%%%%%%%%%%%%%%%%%%%%%%%%%%%%%%%%%%%%%%%%%%%%%%%%%%

%%%%%%%%%%%%%%%%%%%%%%%%%%%%%%%%%%%%%%%%%%%%%%%%%%%%%%%%%%%
%                 TABLE 3
%%%%%%%%%%%%%%%%%%%%%%%%%%%%%%%%%%%%%%%%%%%%%%%%%%%%%%%%%%%
\begin{table}
\caption{Orbital parameters of the elliptic orbits we consider,
plotted for different $\beta$ (colum 1): the eccentricity (column 2),
periastron and apoastron distances (column 3 and 4), orbital period (column
5).  }
\begin{center}
\begin{tabular}{||p{0.08cm}*{6}{c|}|}
\hline
$\beta$ && e & $R_p$ & $R_a$  & T\\[0.5ex]
&& & ($10^4$ km) & ($10^5$ km)  & (s)\\[0.5ex]
\hline
0.4 & &0.25 & 6.0  & 1.0  & 42 \\[0.5ex]
\hline
0.3 & &0.75 & 8.1  & 5.6  & 338 \\[0.5ex]
\hline
0.4 & &0.75 & 6.0  & 4.2  & 220 \\[0.5ex]
\hline
0.5 & &0.75 & 4.8  & 3.4  & 157 \\[0.5ex]
\hline
0.6 & &0.95 & 4.0  & 15.7  & 1337 \\[0.5ex]
\hline
\end{tabular}
\end{center}
\label{table3}
\end{table}
%%%%%%%%%%%%%%%%%%%%%%%%%%%%%%%%%%%%%%%%%%%%%%%%%%%%%%%%%%%
%%%%%%%%%%%%%%%%%%%%%%%%%%%%%%%%%%%%%%%%%%%%%%%%%%%%%%%%%%%
\subsection{Parabolic orbits}
\begin{center}
\begin{figure}
\centerline{\mbox{
\psfig{figure=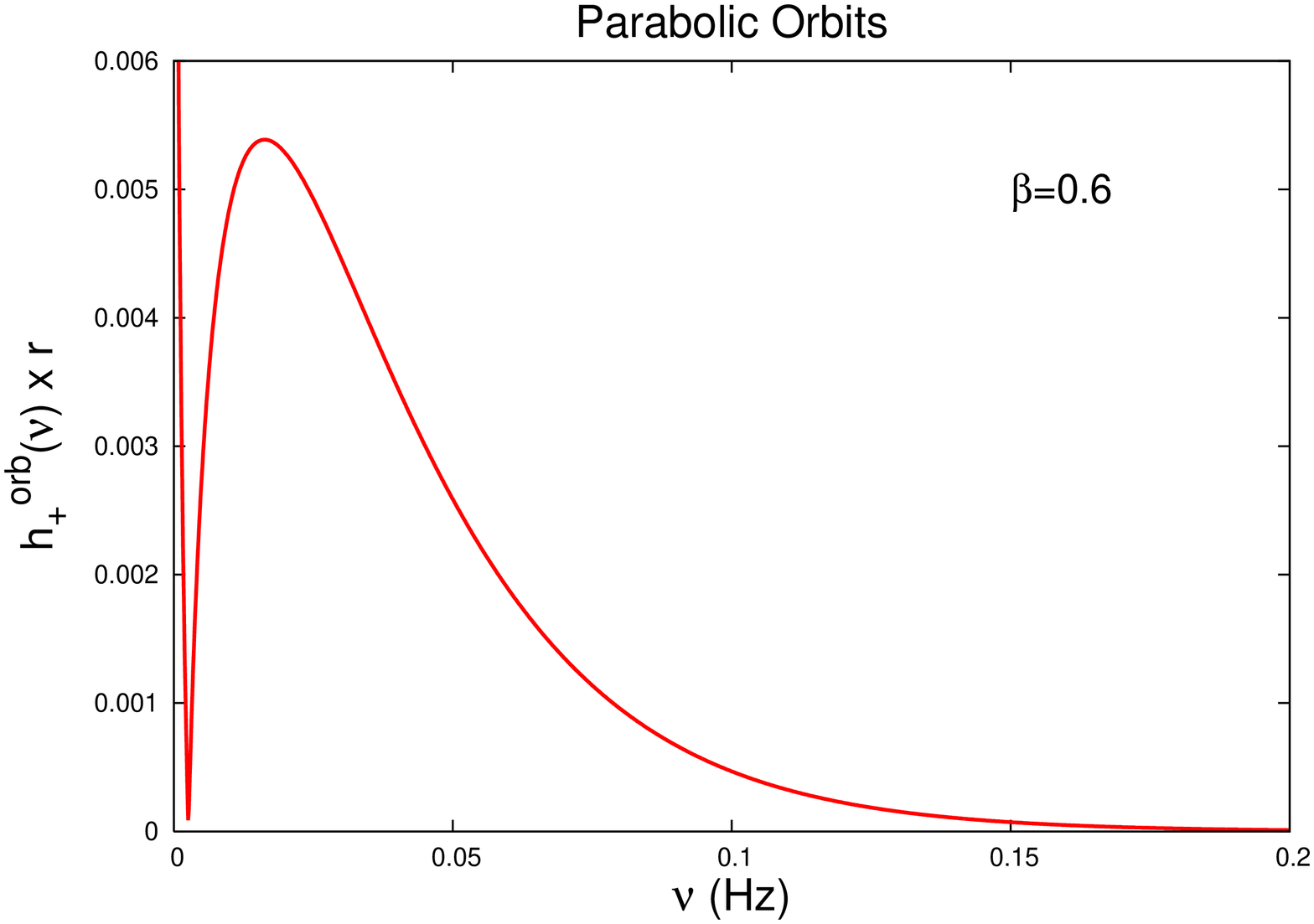,width=9cm,height=7.5cm,angle=-90}~~~~
\psfig{figure=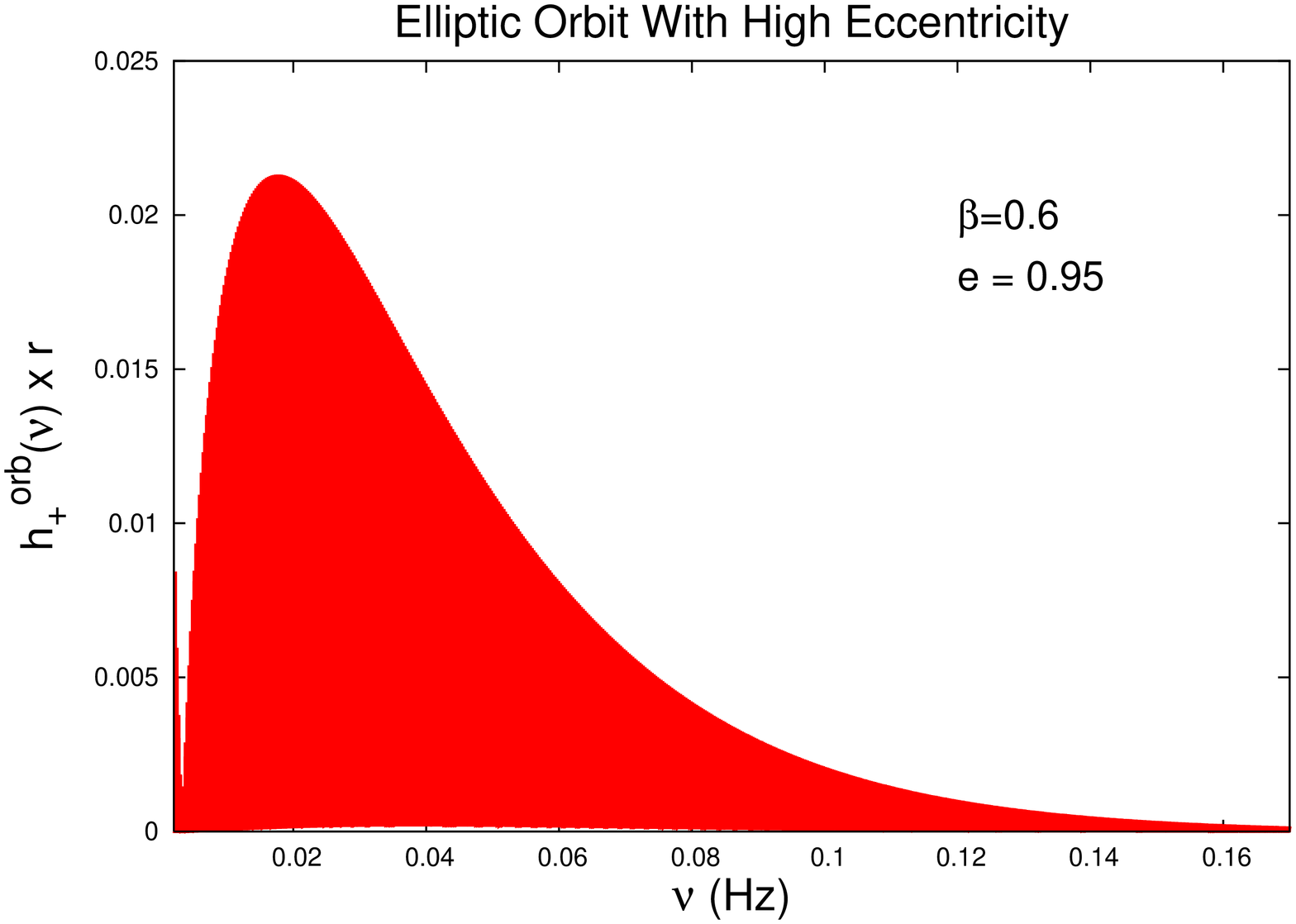,width=9cm,height=7.5cm,angle=-90}
}}
\vskip 12pt
\centerline{\mbox{
\psfig{figure=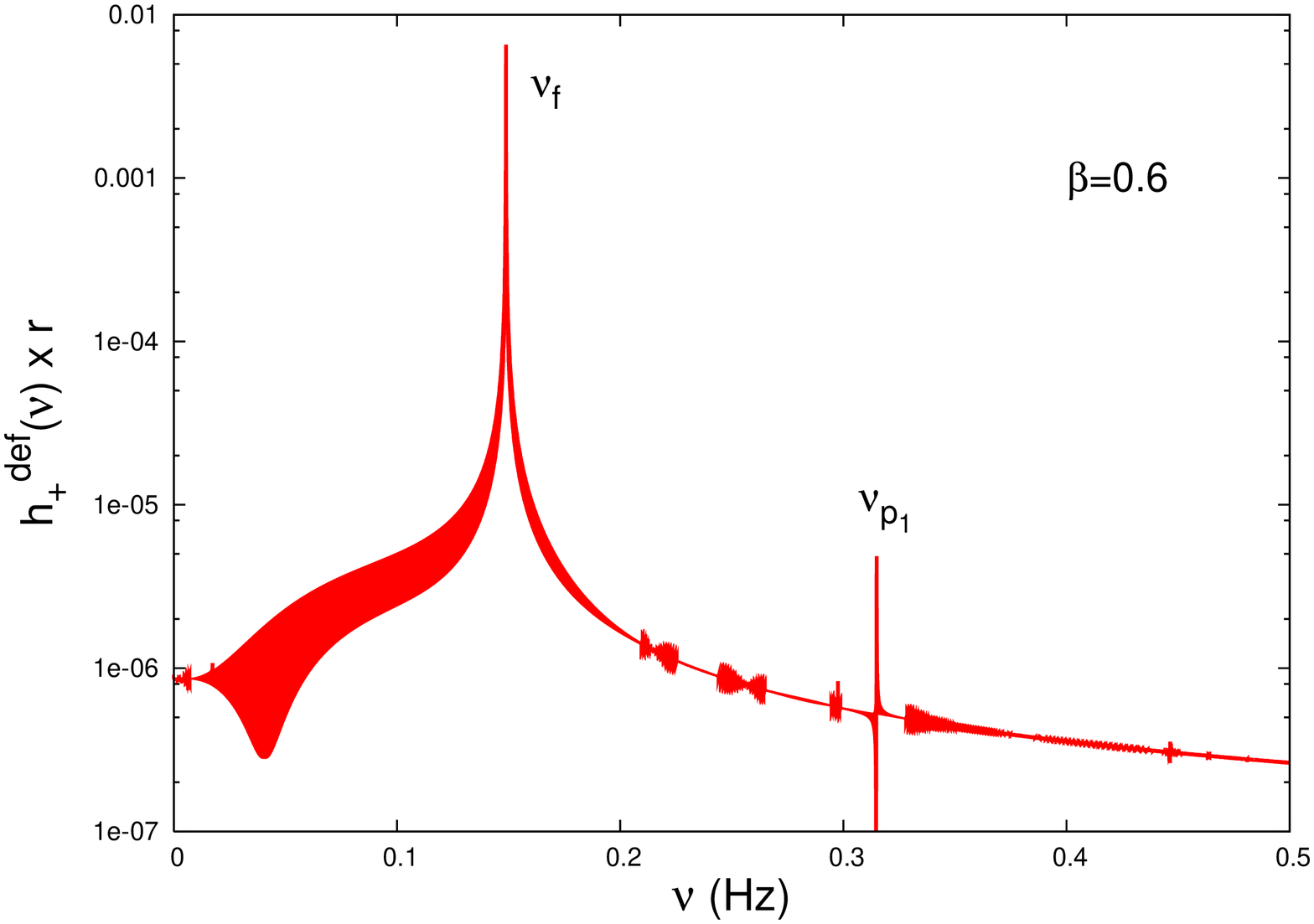,width=9cm,height=7.5cm,angle=-90}~~~~
\psfig{figure=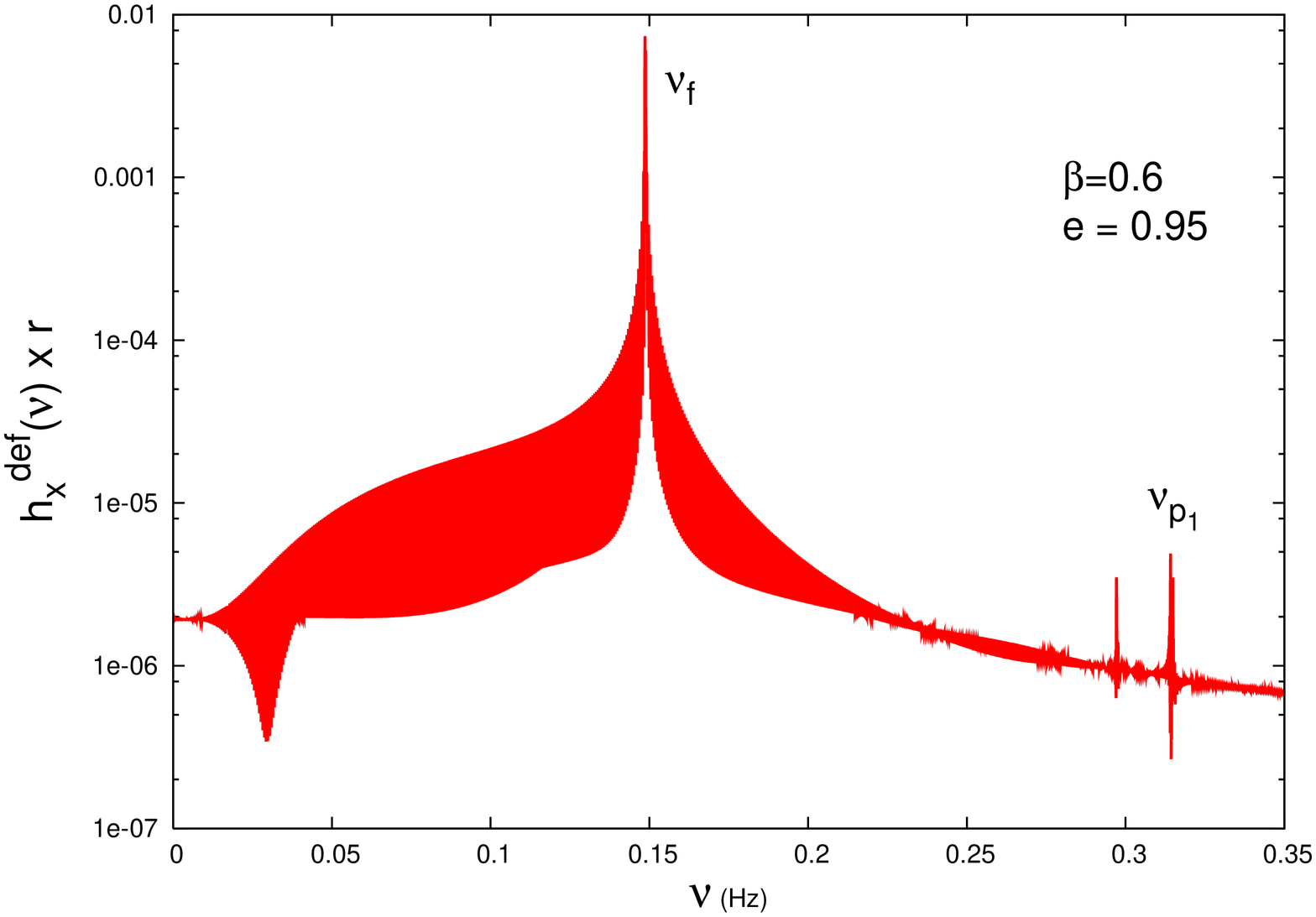,width=9cm,height=7.5cm,angle=-90}
}}
\caption{In the left panel we plot the Fourier transform of the orbital
signal (up) and of the tidal signal (down) emitted when the WD moves on a
parabolic orbit with $\beta=0.6$. For comparison, in the right panel we
show the orbital and the tidal signal when the orbit is elliptic, with the
same $\beta$  and very high eccentricity $e=0.95$.}
\label{FIG5}
\end{figure}
\end{center}

%%%%%%%%%%%%%%%%%%%%%%%%%%%%%%%%%%%%%%%%%%%%%%%%%%%%%%%%%%%
While  for circular and elliptic orbits the orbital GW-signal has
a discrete structure in the frequency domain,
parabolic orbits are associated to signals that are continuous functions of $\nu$.
As an example, in the left panel of figure \ref{FIG5} we show $h^{orb}_+(\nu)$ and
$h^{def}_+(\nu)$ emitted in a parabolic orbit
with $\beta=0.6$.
Again, the mode excitation is manifest in $h^{def}_+(\nu)$, and again the orbital and the
tidal contributions are in a different frequency range; 
but now, unlike the circular and elliptic cases, the dominant contribution
appears to be the tidal one, reaching a peak-amplitude of about $\sim 10^{-2}$ at
$\nu_f=0.146$ Hz, about twice the amplitude of the orbital peak, which occurs at $\nu= 1.6\cdot
10^{-2}$ Hz. 
%%%%%%%%%%%%%%%%%%%%%%%%%%%%%%%%%%%%%%%%%%%%%%%%%%%%%%%%%%%%
%                      FIGURE
%%%%%%%%%%%%%%%%%%%%%%%%%%%%%%%%%%%%%%%%%%%%%%%%%%%%%%%%%%%%
%%%%%%%%%%%%%%%%%%%%%%%%%%%%%%%%
%%%%%%%%%%%%%%%%%%%%%%%%%%%%%%%
Comparing the orbital signal of figure \ref{FIG5}  with the orbital signal in figure  
\ref{FIG4} (upper panel, left)
we see that the shape of the parabolic signal is basically the envelope of the elliptic one;
indeed, the parabola can be seen as the limiting case of an ellipse with eccentricity equal to
unity.
It is interesting to compare the signal emitted on the parabolic orbit
with that  emitted on a very elongated elliptic orbit with the same $\beta$:
in the right panel of figure \ref{FIG5} we show
the orbital (up) and the tidal (down) signals emitted when the orbit is an
ellipse with $\beta=0.6$ and $e=0.95$.
We see that in this case the amplitude of
the {\bf f}-mode peak increases significantly with respect to the
orbital one, though it does not become bigger. Thus it appears that, for a fixed $\beta$,
the oscillations modes of
the star increasingly contribute to the total emitted radiation
as the orbit elongation increases,
and become dominant when the orbit is a parabola.

\begin{center}
\begin{figure}
\centerline{\mbox{
\psfig{figure=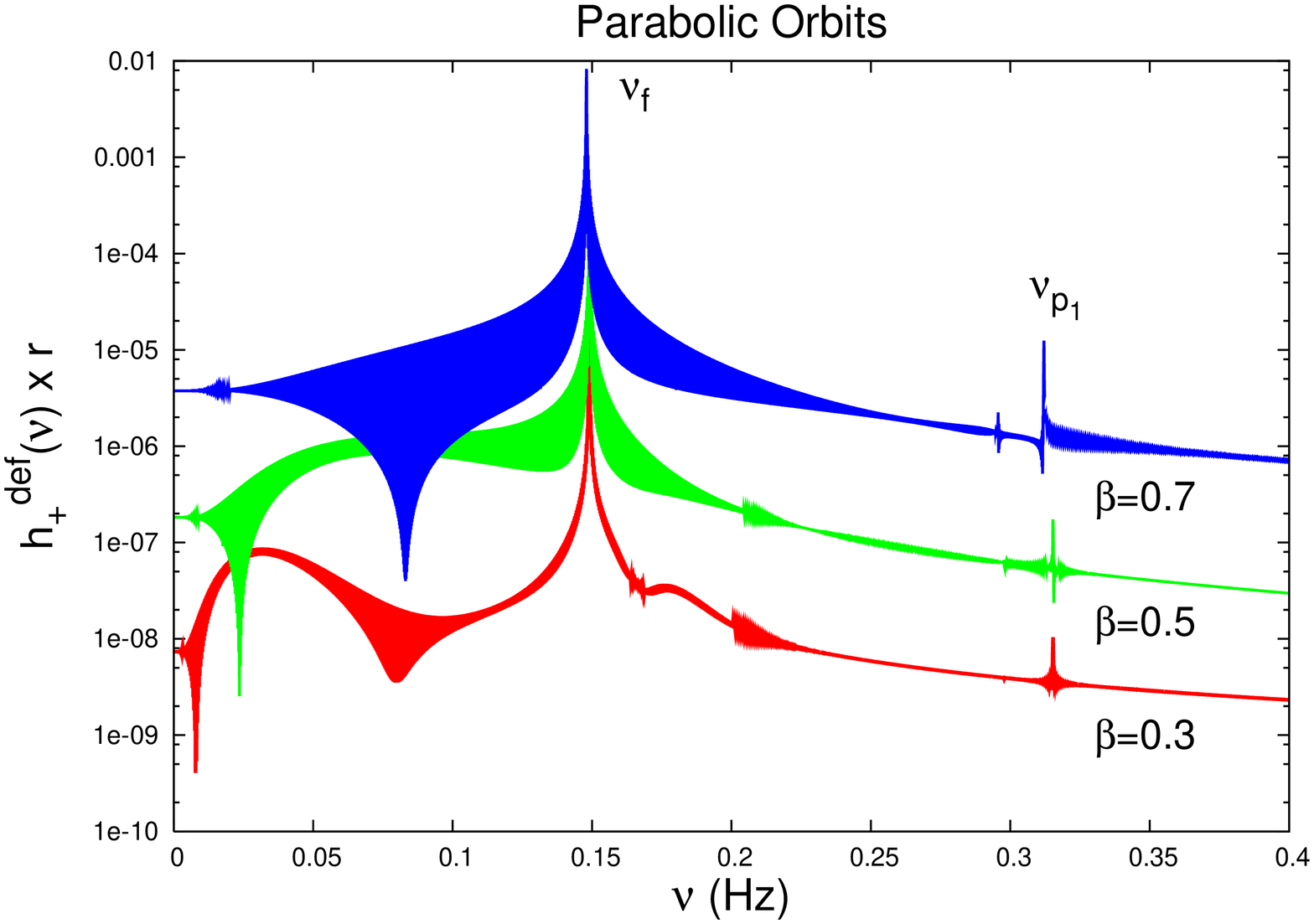,width=9cm,height=7.5cm,angle=-90}
}}
\caption{The Fourier transform of the tidal signal emitted in parabolic
orbits is plotted as a function of frequency for three values of $\beta$.}
\label{FIG6}
\end{figure}
\end{center}

In figure \ref{FIG6} we plot the tidal signal for parabolic orbits
for  different values of the penetration factor.
It should be mentioned that, as for circular and elliptic orbits, the signals emitted 
along the axis orthogonal to the orbital plane in the
'$\times$'-polarization have a structure  similar to that of the '+'- polarization,
and for this reason we omit to show them.
%%%%%%%%%%%%%%%%%%%%%%%%%%%%%%%%%%%%%%%%%%%%%%%%%%%%%%%%%%%%
%                      FIGURE
%%%%%%%%%%%%%%%%%%%%%%%%%%%%%%%%%%%%%%%%%%%%%%%%%%%%%%%%%%%%
%%%%%%%%%%%%%%%%%%%%%%%%%%%%%%%
%%%%%%%%%%%%%%%%%%%%%%%%%%%%%%%
As a last example, we have computed the radiation emitted when the star penetrates 
more deeply into the black hole tidal radius, choosing values of $\beta> 1$, but always
lower than the tidal disruption limit. In this case the star is highly deformed and assumes
a `cigar' shape; therefore, we cannot expect to see in the frequency spectrum
the peaks corresponding to the
excitation of the modes of the spherical star, as it was in previous cases.
%%%%%%%%%%%%%%%%%%%%%%%%%%%%%%%%%%%%%%%%%%%%%%%%%%%%%%%%%%%%
%                      FIGURE
%%%%%%%%%%%%%%%%%%%%%%%%%%%%%%%%%%%%%%%%%%%%%%%%%%%%%%%%%%%%
\begin{center}
\begin{figure}
\centerline{\mbox{
\psfig{figure=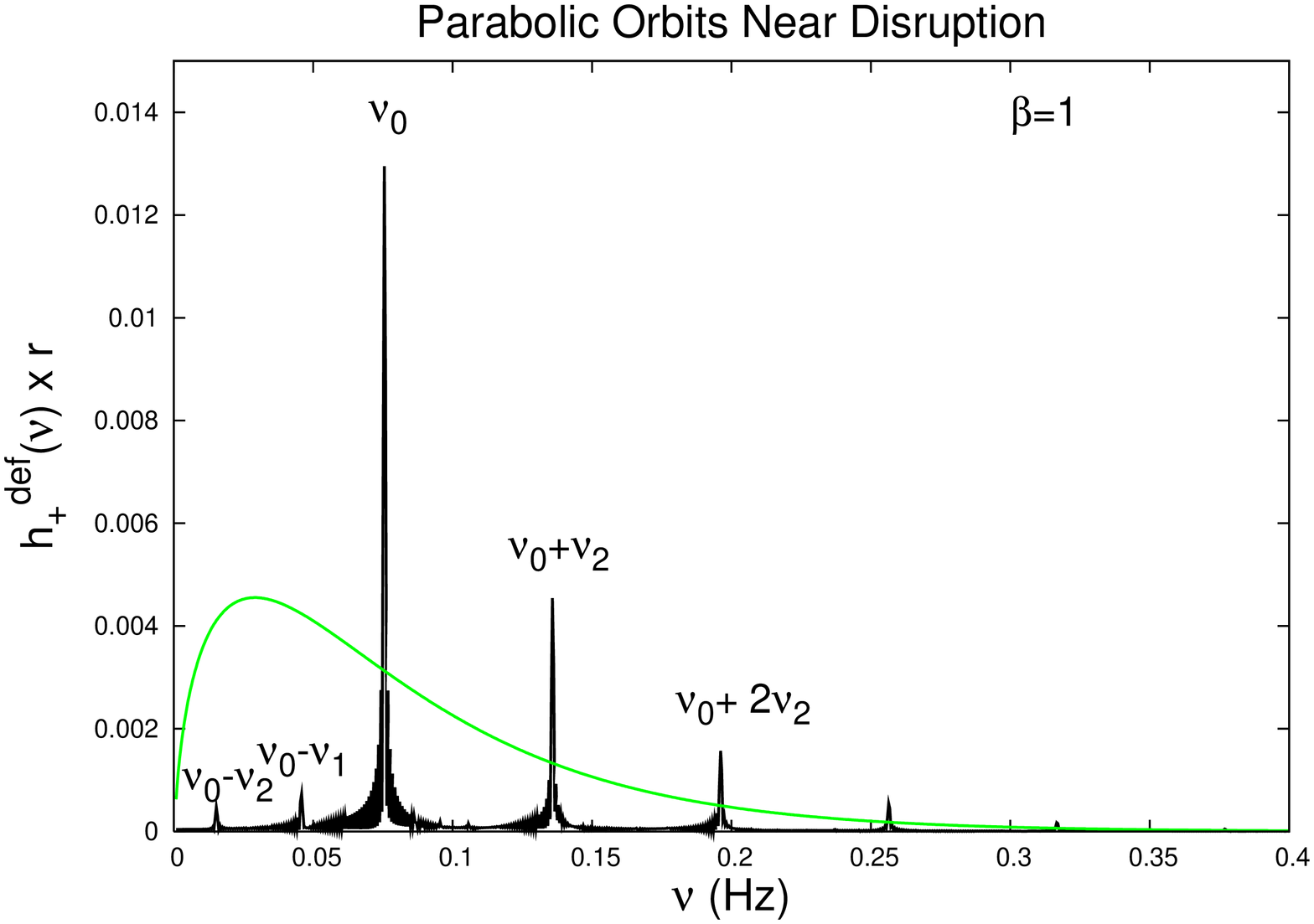,width=9cm,height=7.3cm,angle=-90}~~~~
\psfig{figure=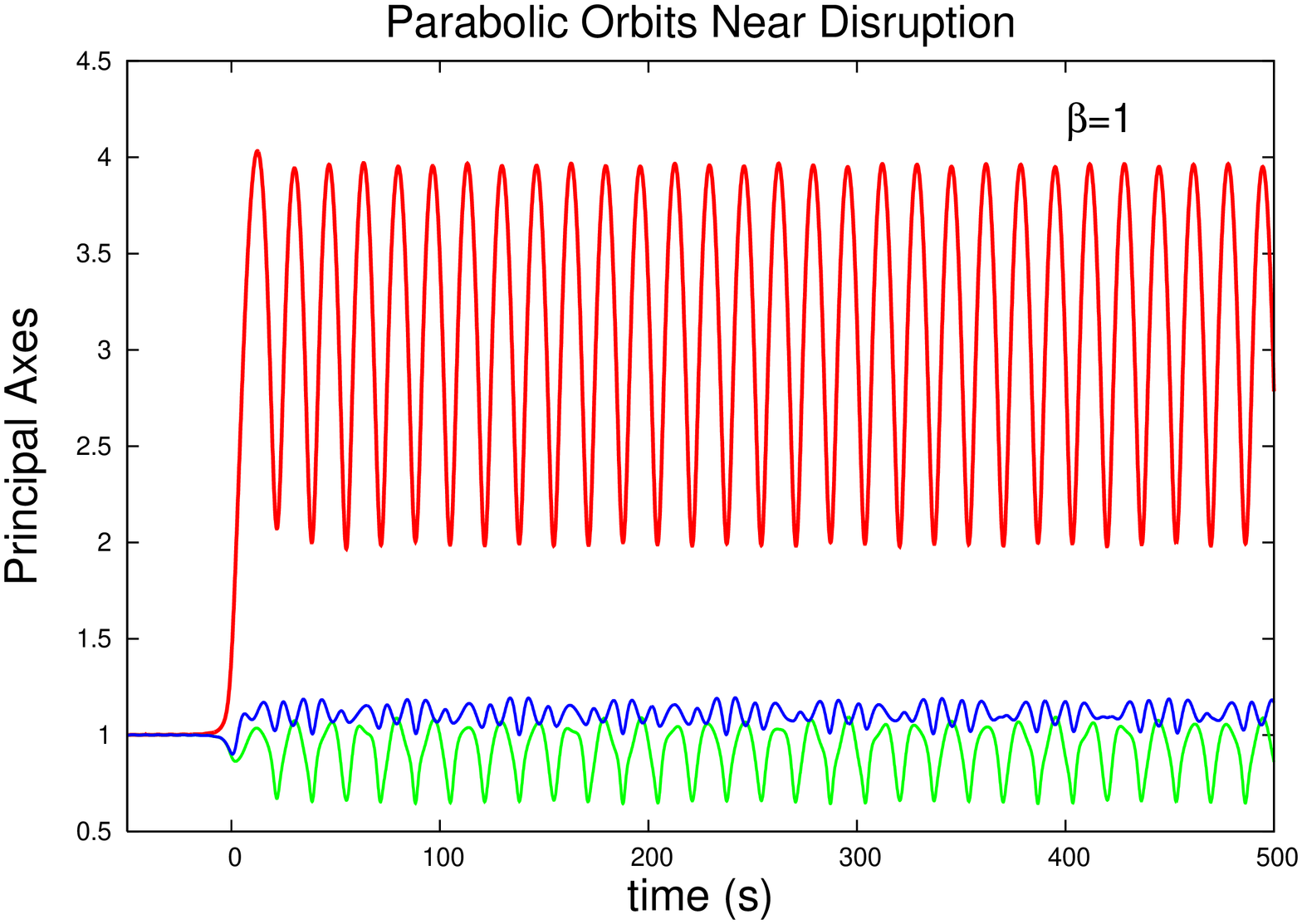,width=9cm,height=7.3cm,angle=-90}
}}
\vskip 12pt
\centerline{\mbox{
\psfig{figure=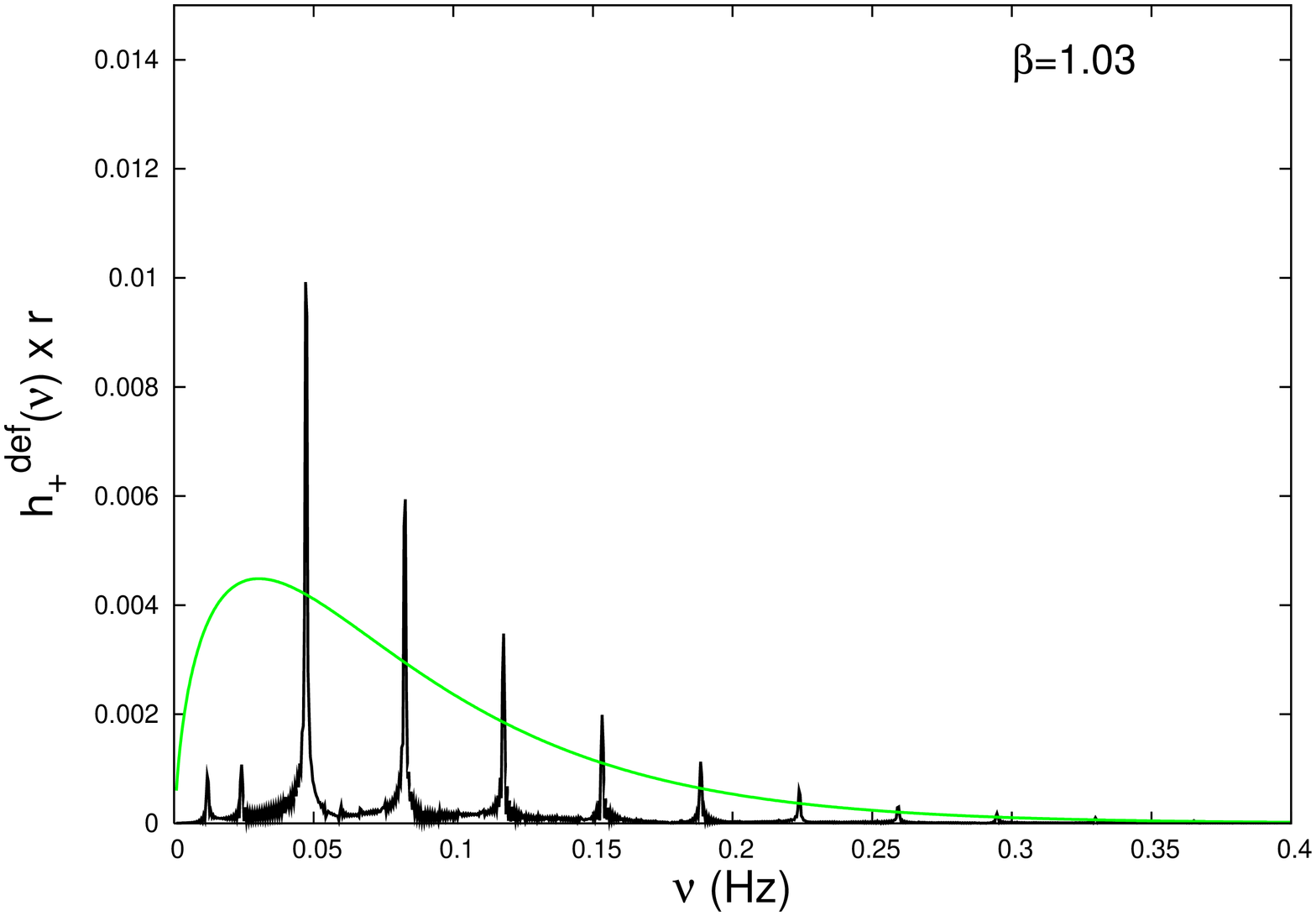,width=9cm,height=7.3cm,angle=-90}~~~~
\psfig{figure=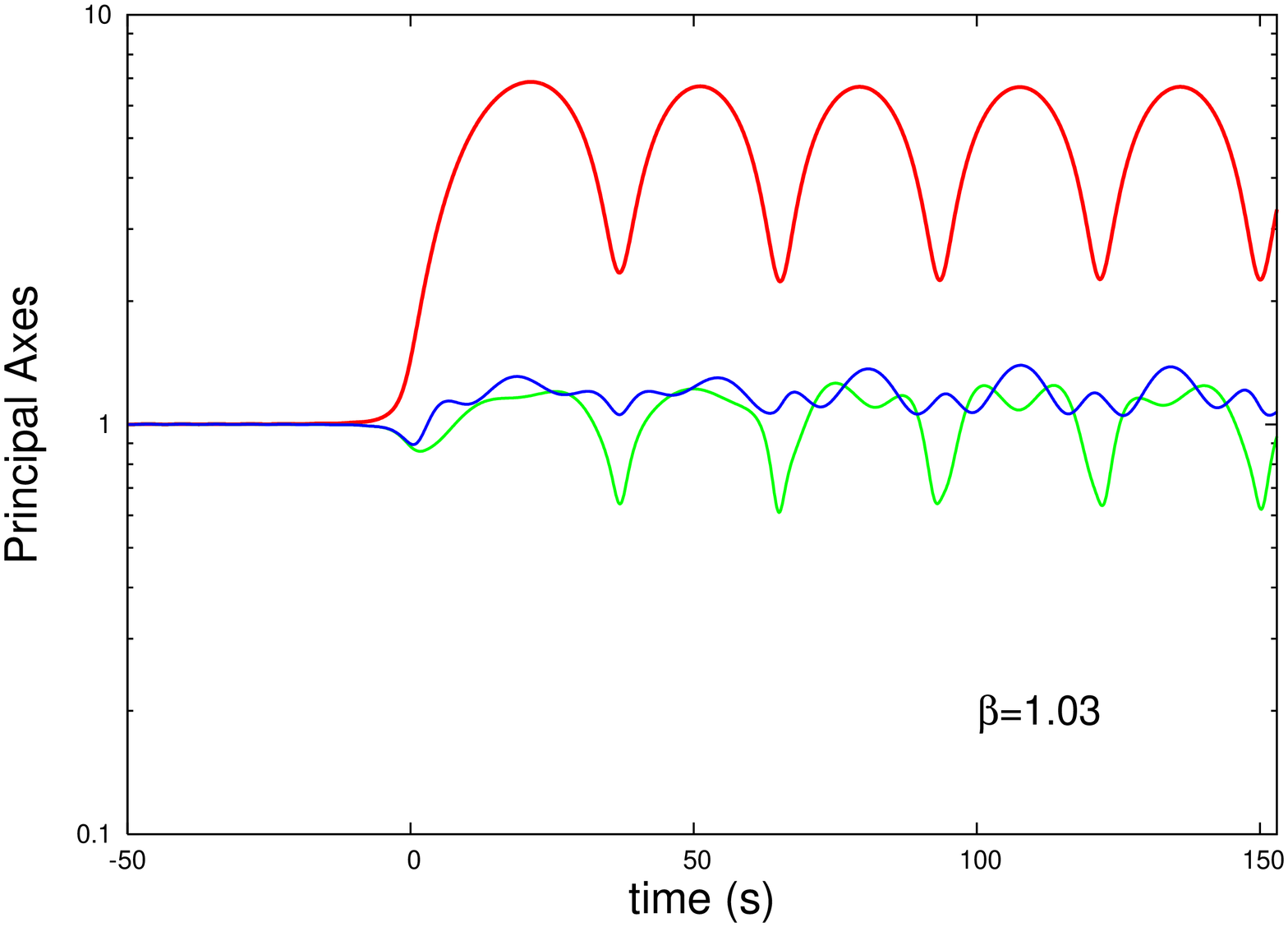,width=9cm,height=7.3cm,angle=-90}
}}
\vskip 12pt
\centerline{\mbox{
\psfig{figure=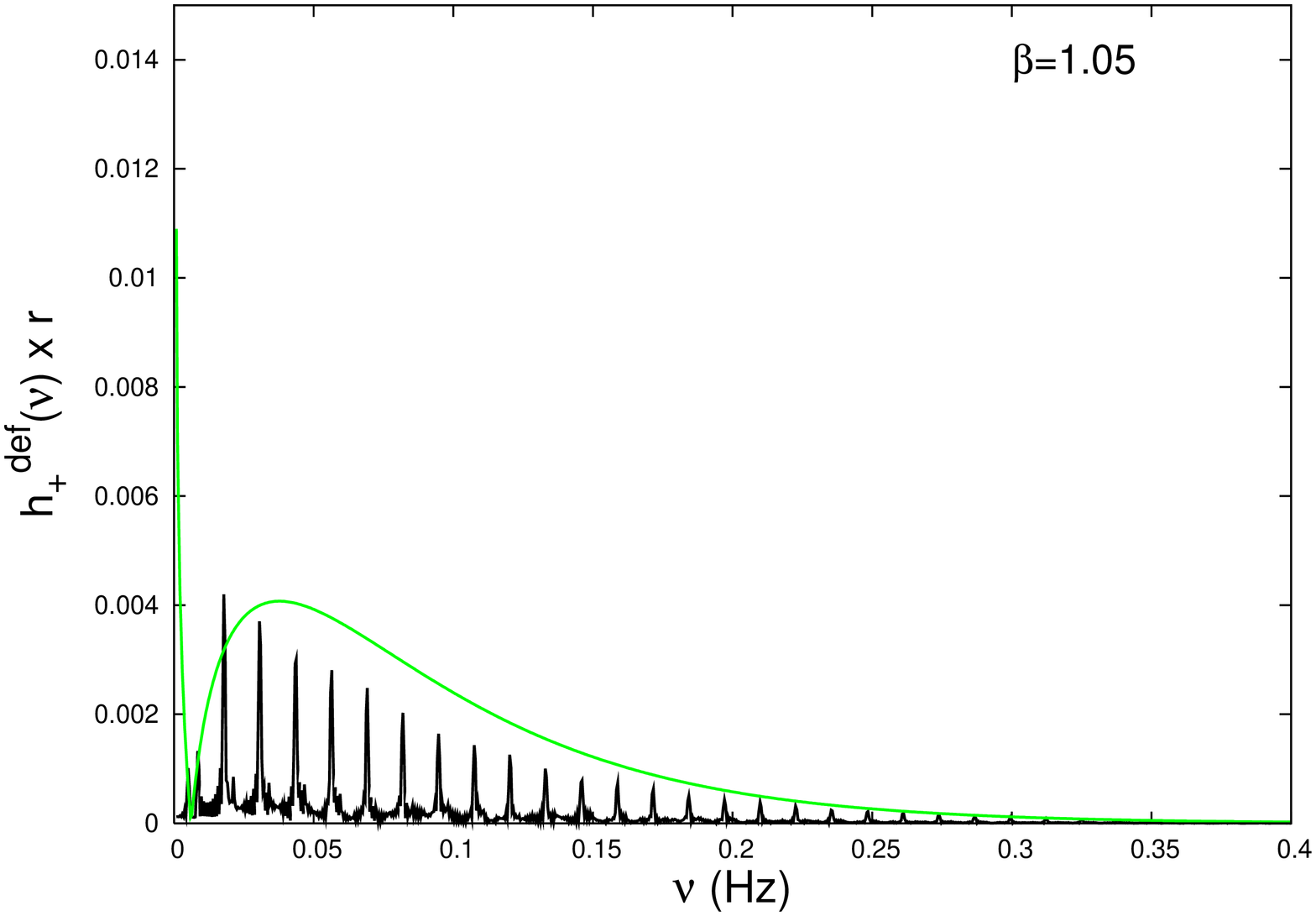,width=9cm,height=7.3cm,angle=-90}~~~~
\psfig{figure=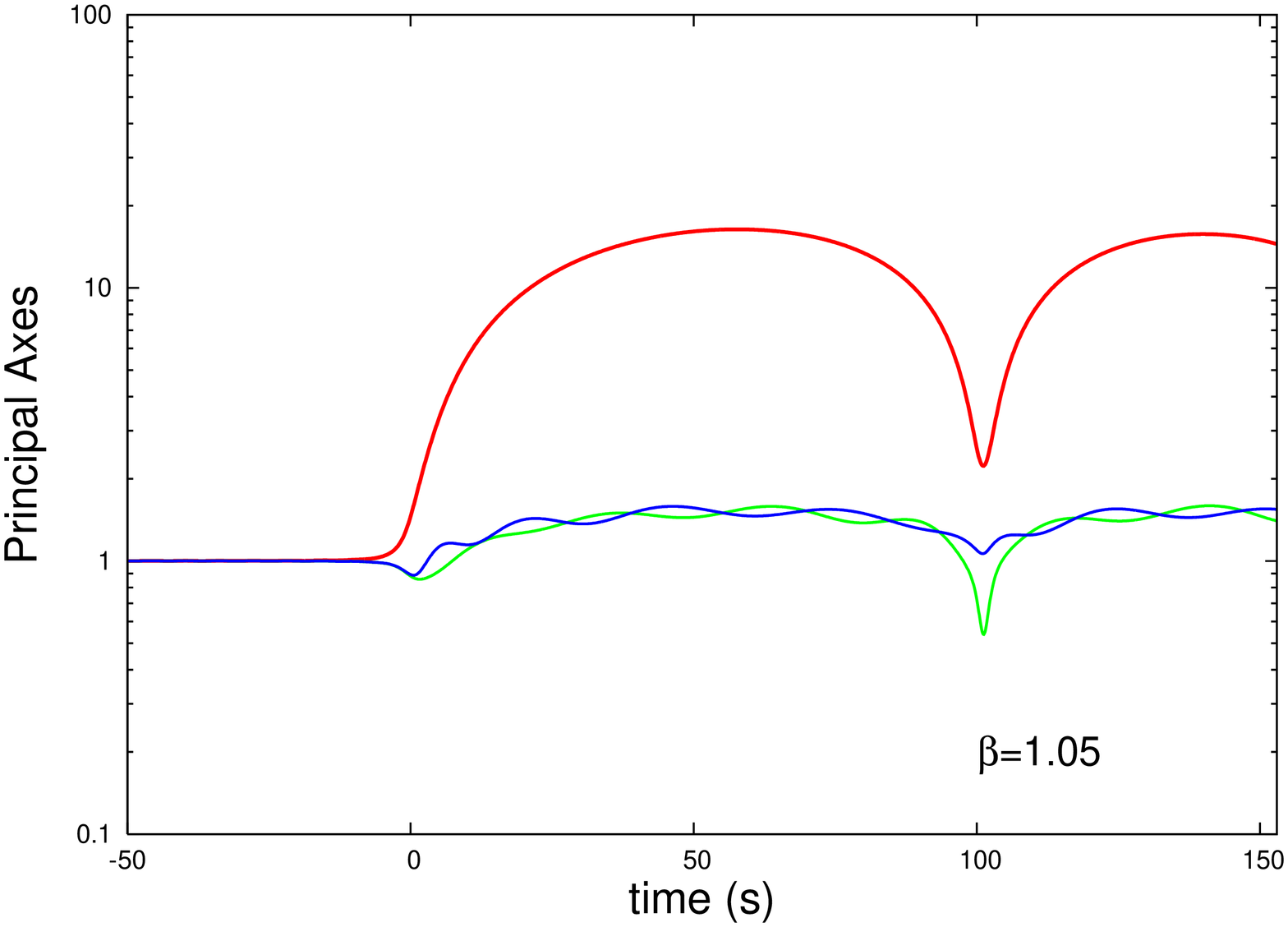,width=9cm,height=7.3cm,angle=-90}
}}
\caption{The tidal signal emitted in a parabolic orbit is plotted for three values
of $\beta$ approaching the critical value (left panel). The green continuous
line is the orbital signal, plotted for comparison.
On the right panel we show the behaviour of the principal axes of the star as a function of
time, when the star passes through the periastron.
The relation between the axes behaviour and  the emitted signal,  and the origin of
different peaks in $h^{def}$ are explained in the text.}
\label{FIG7}
\end{figure}
\end{center}
%%%%%%%%%%%%%%%%%%%%%%%%%%%%%%%
%%%%%%%%%%%%%%%%%%%%%%%%%%%%%%%
However, the cigar also oscillates and his eingenfrequencies scale
as the inverse of his length, being the mass fixed. 
Indeed the tidal signal presents a very interesting structure shown in figure \ref{FIG7}.
In the three panels on the left we show  $h^{def}_+(\nu)$ for $\beta=1, ~1.03,~1.05$.
We see that there is a dominant peak which
corresponds to the excitation of the fundamental mode of the cigar; indeed, as 
expected, when $\beta$ increases  the tidal interaction is stronger, 
the star  assumes a more elongated cigar shape and the frequency of the main peak decreases.
The spectrum also shows  several equally spaced spectral lines and 
their origin can be understood by looking at the right panel of  figure
\ref{FIG7}, were we
plot the principal axes of the star as functions of time. 
When the star approaches the black hole, one of the axes on the equatorial plane -
let us name it `a'- grows much more than the other two, `b' and `c'.
In the figure the axes are normalized to their initial values, therefore 
at large distance (large negative time) they are all equal to 1 since the star is spherical. 
Analysing the axes behaviour, we can identify two characteristic frequencies: one, $\nu_1$, which corresponds to the first 
oscillation,  when the a-axis  starts to grow, reaches the maximum elongation, and then
decreases to the value about which it will oscillate in the new, approximately Riemaniann,
stationary configuration; 
the second one, $\nu_2,$ corresponds to the oscillation of the a-axis about the new
configuration. For the three values of $\beta$ considered in figure
\ref{FIG7} these frequencies are given in table \ref{table4}. 
Referring, for example, to the signal $h^{def}_+(\nu)$ for $\beta=1$ in the left panel of figure
\ref{FIG7}, and indicating with $\nu_0$ the frequency corresponding  to the fundamental mode of
the `cigar', we see that there are equally spaced  spectral lines at
$\nu= \nu_{0} \pm n \nu_2,$ with $n=1,2,...$, and smaller peaks (only the first 
on the left of the main peak is clearly visible) which correspond to $\nu= \nu_{0} \pm n
\nu_1$.
The spacing between the peaks corresponds exactely to the frequencies, given in table
\ref{table4}, at which the axes oscillate. 
%%%%%%%%%%%%%%%%%%%%%%%%%%%%%%%%%%%%%%%%%%%%%%%%%%%%%%%%%%%
%              TABLE 4
%%%%%%%%%%%%%%%%%%%%%%%%%%%%%%%%%%%%%%%%%%%%%%%%%%%%%%%%%%%
\begin{table}
\caption{In this table we give, for three values of $\beta$ approaching the critical value
and for parabolic orbits, three  frequencies characteristic of the tidal deformation:
$\nu_0$, is the frequency of the main peak appearing in the tidal signals
shown in figure \ref{FIG7} and
corresponds to the excitation of the fundamental mode of the deformed star;
$\nu_1$, refers to the  first oscillation of the a-axis 
as the star approaches the periastron, and 
$\nu_2$ corresponds to the oscillation of the a-axis about the new
cigar-like configuration. }
\begin{center}
\begin{tabular}{||p{0.08cm}*{4}{c|}|}
\hline
$\beta$ & $\nu_{0}$ &  $\nu_{1}$  &  $\nu_{2}$ \\[0.5ex]
 & ($10^{-2}$ Hz)  & ($10^{-2}$ Hz) & ($10^{-2}$ Hz) \\[0.5ex]
\hline
1 & 7.5 &4.6& 6.0  \\[0.5ex]
\hline
1.03 & 4.7 & 2.5& 3.5 \\[0.5ex]
\hline
1.05 & 1.8 &0.8&  1.3  \\[0.5ex]
\hline
\end{tabular}
\end{center}
\label{table4}
\end{table}
%%%%%%%%%%%%%%%%%%%%%%%%%%%%%%%%%%%%%%%%%%%%%%%%%%%%%%%%%%%

%%%%%%%%%%%%%%%%%%%%%%%%%%%%%%%%%%%%%%%%%%%%%%%%%%%%%%%%%%%%
\subsection{Results for a different black hole mass}
%%%%%%%%%%%%%%%%%%%%%%%%%%%%%%%%%%%%%%%%%%%%%%%%%%%%%%%%%%%%

It is interesting to see how the entire  picture discussed 
above changes if we change the black hole mass.
Be $R_p$ the periastron distance (coincident with the orbital radius in the
circular case).
The tidal tensor (\ref{tidal_tensor_newton}) scales as
$C_{ij} \propto (M_{BH}/R_p^3) =\beta^3$; therefore we expect that,
for an assigned $\beta$, both  the tidal signal $h^{def}$ 
and the critical value  of $\beta$ needed for disruption are independent of the black hole mass.
Indeed, this is what we find by numerical integration. 

\begin{center}
\begin{figure}
\centerline{\mbox{
\psfig{figure=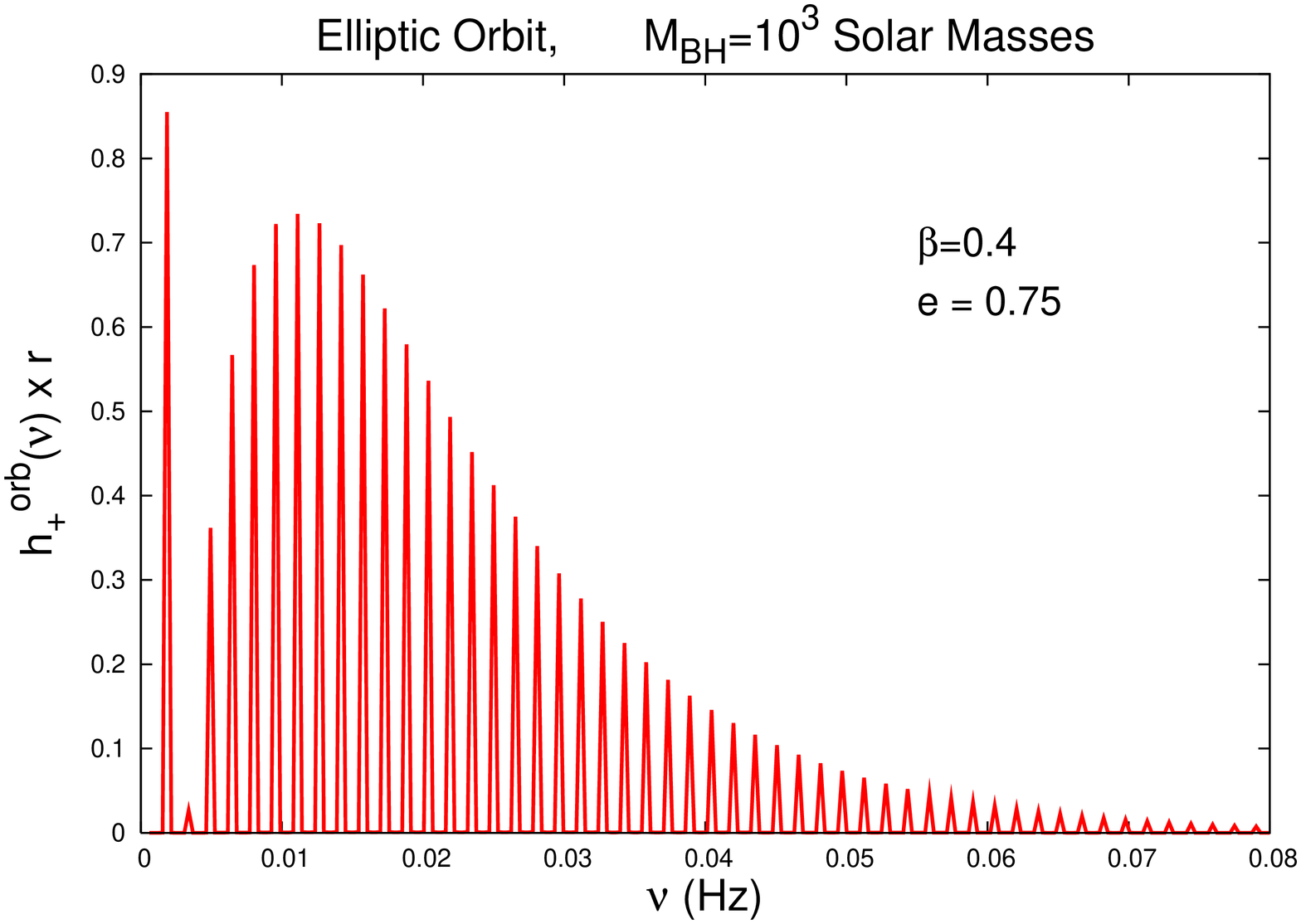,width=9cm,height=7.5cm,angle=-90}~~~~
\psfig{figure=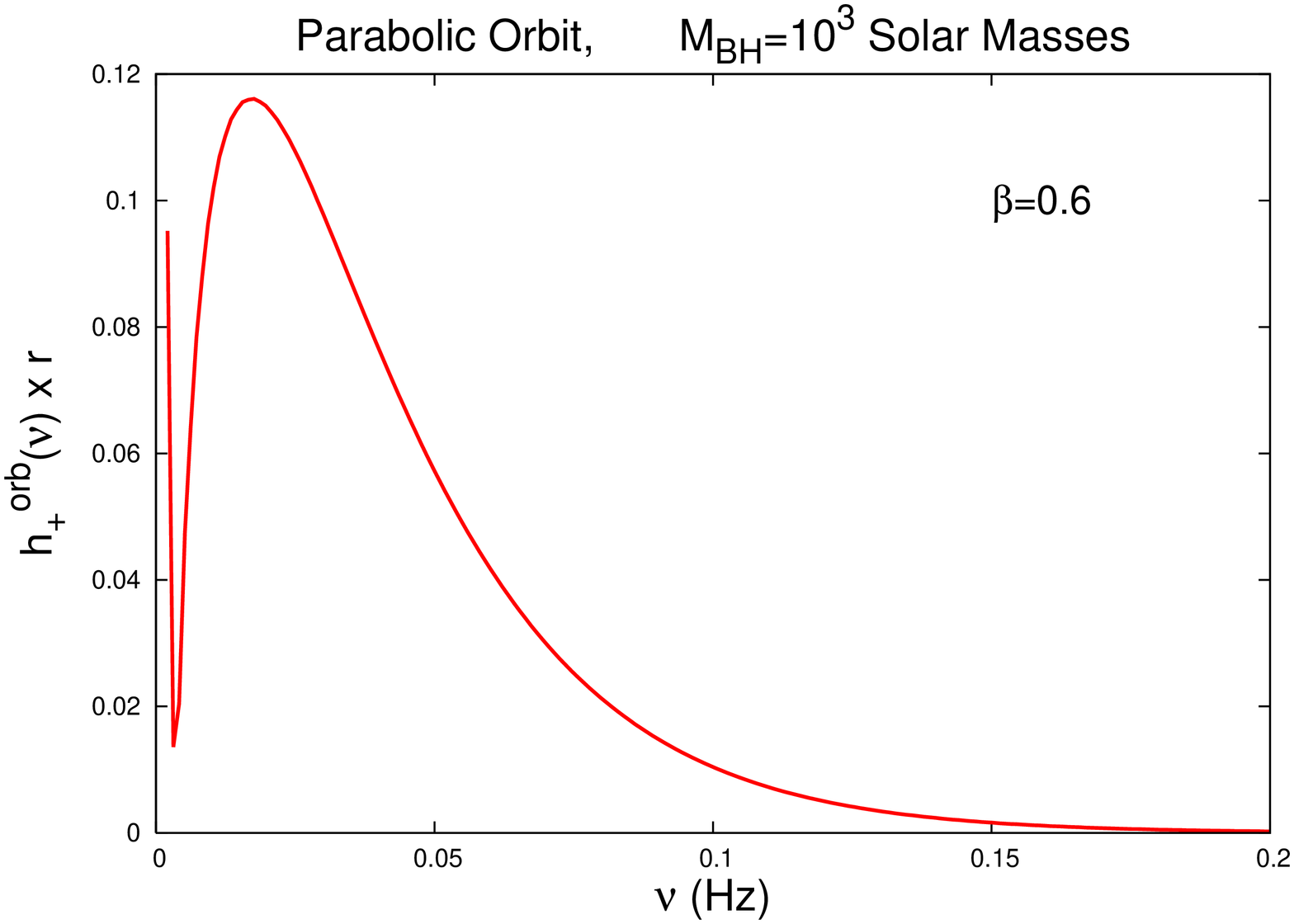,width=9cm,height=7.5cm,angle=-90}
}}
\caption{The orbital signal emitted  when the black hole mass is $M_{BH}=10^3 \msun$ is plotted
for an eccentric orbit with $\beta=0.6$ and $e=0.75$ (left)
and for a parabolic orbit with $\beta=0.6$ (right).
These plots have to be compared to the upper panel of figure \ref{FIG4} 
and \ref{FIG5}, respectively, where the orbital signal is plotted for the same orbital
parameters and a black hole mass $M_{BH}= 10 \msun$. }
\label{FIG8}
\end{figure}
\end{center}

Conversely, the orbital part of the signal, $h^{orb},$ depends on the black hole mass.
For instance, in the circular case from eq. (\ref{amplcirc}) we see that for $M_* << M_{BH}$
the amplitude of the emitted spectral line is $A \propto M_{BH}$, showing that the orbital signal increases 
with $M_{BH}$.   A similar behaviour is exhibited in  elliptic and parabolic orbit, as shown in
figure \ref{FIG8}, where we plot $h_+^{orb}$ emitted  when 
$M_{BH}=10^3 \msun,$ for an elliptic orbit with $\beta=0.6$ and $e=0.75$ (left)
and for a parabolic orbit with $\beta=0.6$ (right).
A comparison of these signals with those found for the same parameters when the black hole mass
was $M_{BH}=10 \msun$ (upper panel, left, of figure \ref{FIG4} and \ref{FIG5}, respectively)
clearly shows how the orbital signal increases with $M_{BH}$.
%%%%%%%%%%%%%%%%%%%%%%%%%%%%%%%%%%%%%%%%%%%%%%%%%%%%%%%%%%%%
%                      FIGURE
%%%%%%%%%%%%%%%%%%%%%%%%%%%%%%%%%%%%%%%%%%%%%%%%%%%%%%%%%%%%
%%%%%%%%%%%%%%%%%%%%%%%%%%%%%%%
%%%%%%%%%%%%%%%%%%%%%%%%%%%%%%%

%%%%%%%%%%%%%%%%%%%%%%%%%%%%%%%%%%%%%%%%%%%%%%%%%%%%%%%%%%%
\section{Concluding Remarks}
%%%%%%%%%%%%%%%%%%%%%%%%%%%%%%%%%%%%%%%%%%%%%%%%%%%%%%%%%%%
In this paper we have computed the gravitational signal emitted when a white dwarf 
moves around a black hole on a closed or open orbit. We have computed both the orbital and the
tidal contributions and compared the two, assuming that the star is close to the
black hole, but in a region safe enough to prevent its tidal disruption.
In all cases, the non radial oscillation modes of the star are excited (in our approach we do
not include the dynamical behaviour of the black hole)
to an extent which depends on how deep in the tidal radius the star penetrates and on the
type of orbit.
The orbital, $h^{orb}$,  and the tidal, $h^{def}$, contributions are emitted 
at different frequencies: smaller for $h^{orb}$, higher for  $h^{def}$.

The system we have analyzed in detail is composed of a $1~\msun$ white dwarf and a
$10~\msun$ black hole. In this case we find that for circular orbits up to the critical one,
the amplitude of the single spectral line in $h^{orb}$ is always smaller than that of
the {\bf f}-mode peak in $h^{def}$; however if the black hole mass is larger,
the situation reverses, since $h^{def}$ is nearly independent of $M_{BH}$, while
$h^{orb}$ is proportional to it.
It is interesting to note that $h^{def}$ contains several peaks that shows the coupling between
the orbital and tidal motion.

When the orbit is elliptic, $h^{orb}$ shows several spectral lines emitted at multiples of the
orbital frequency; $h^{def}$ has again a main peak at the {\bf f}-mode frequency,  
and in addition a number of peaks due to the orbital-tidal interaction, that contain a signature
of the nature of the orbit. 
The amplitude of the {\bf f}-mode peak of course increases with the penetration factor $\beta$
and also, for a fixed $\beta$, with the eccentricity;  for  the system we have considered 
this amplitude tends to that of the maximum of $h^{orb}$ for very elongated orbits ($e > 0.95$).
Again changing the black hole mass to a higher value the orbital contribution increases with
respect to the tidal one.

For parabolic orbits the situation is very intersting, since the star can penetrate more deeply
into the tidal radius, allowing for larger values of $\beta$.
As $\beta$ tends to the critical value the star assumes a very elongated, cigar-like shape,
deviating considerably by its initial spherical structure. Thus the main peak in $h^{def}$
shifts toward lower frequencies, since the {\bf f}-mode frequency scales as the inverse of the
lenght of the principal axis. The harmonics which appear in the tidal signal 
corresepond to the oscillation frequencies of
the  principal axis of the `cigar', and the larger is $\beta$ the higher will be the number of
excited harmonics. A comparison with the behaviour of the orbital signal (see figure
\ref{FIG7}) shows that if $\beta < 1.03$ 
the tidal signal is considerably higher than  $h^{orb}$
if $M_{BH}= 10~\msun$, but again the situation reverses if the black hole mass is sufficiently
higher.

The results of our study show that the tidal signals emitted by a close encounter
between a white dwarf and a black hole lay in a frequency region  which is intermediate
between the sensitivity region of LISA ($10^{-4}-10^{-1}$ Hz) and that of ground-based
intereferometers VIRGO, LIGO, GEO, TAMA, sensitive above $\sim 10$ Hz.
Detectors that fill this gap have recently been proposed,
like  the Big Bang Observatory, thought as follow on mission to LISA \cite{bender_etal_2005},
and DECIGO, proposed by a Japaneese group  \cite{Seto_Kawamura_Nakamura_2001};
they should both be extremely sensitive in the decihertz region  $10^{-1}-1$ Hz,
and would be the appropriate instruments to detect the tidal signals we have studied
and to shed light on the dynamics of dense stellar clusters where these 
processes are more likely to occur.

%%%%%%%%%%%%%%%%%%%%%%%%%%%%%%%%%%%%%%%%%%%%%%%%%%%%%%%%%%%
\begin{center}
\bf{Acknowledgements}
\end{center}

We would like to thank L. Rezzolla for introducing us to the literature on the subject,
and L. Gualtieri for useful conversations and fruitful insights in various aspects
of the work.

%%%%%%%%%%%%%%%%%%%%%%%%%%%%%%%%%%%%%%%%%%%%%%%%%%%%%%%%%%%%%%%%%%
%                      REFERENCES
%%%%%%%%%%%%%%%%%%%%%%%%%%%%%%%%%%%%%%%%%%%%%%%%%%%%%%%%%%%%%%%%%%

\end{document}